\documentclass[twocolumn,           % Format : preprint, twocolumn
               showpacs,            % Pacs : showpacs, noshowpacs
               preprintnumbers,     % Preprint: preprintnumbers,
               aps,                 % Society: ...
               prd,                 % Journal Style : pra, prb, prc, prd, pre,
               a4paper,             % Size : a4paper, ...
               superscriptaddress,  % Affiliation (Title) : groupedaddress,
               nofootinbib,         % Footnote: footinbib, nofootinbib
               tightenlines,        % Remove additional spaces in a line
               floats,floatfix      % Floating pictures and tables
               ]{revtex4}

\pdfoutput=1
\usepackage{ graphicx }
\usepackage{ latexsym }
\usepackage{ amsmath, amssymb }        % amssymb includes amsfonts
\usepackage[ draft=false ]{ hyperref }
\usepackage{multirow}
\usepackage{longtable}

\newcommand{\rd}{\mathrm{d}}
\newcommand{\DLie}{\mathcal{L}}
\newcommand{\R}{\mathbb{R}}
\newcommand{\e}{\mathrm{e}}

\newcommand{\MADM}{M_{\mathrm{ADM}}}
\newcommand{\fR}{$f(R)$ }
\newcommand{\Olliptic}{\textsc{Olliptic} }
\newcommand{\AMSSNCKU}{\textsc{AMSS-NCKU} }
\newcommand{\bfa}{\textbf{(a)}}
\newcommand{\bfb}{\textbf{(b)}}
\newcommand{\bfc}{\textbf{(c)}}
\newcommand{\bfd}{\textbf{(d)}}
\newcommand{\imgdir}{.}

\usepackage{color}

\begin{document}

% ----> TITLE PAGE

%|--------------------------------------------------------------------|
\title{Binary Black Hole merger in \fR theory}
%|--------------------------------------------------------------------|

%|--------------------------------------------------------------------|
\author{Zhoujian Cao}
\email{zjcao@amt.ac.cn}
\affiliation{Institute of Applied  Mathematics, Academy of Mathematics
  and Systems  Science, Chinese  Academy of Sciences,  Beijing 100190,
  China}
%|--------------------------------------------------------------------|

%|--------------------------------------------------------------------|
\author{Pablo Galaviz}
\email{Pablo.Galaviz@monash.edu}
\affiliation{School   of  Mathematical  Science,   Monash  University,
  Melbourne, VIC 3800, Australia}
%|--------------------------------------------------------------------|

%|--------------------------------------------------------------------|
\author{Li-Fang Li}
\email{lilf@itp.ac.cn}
\affiliation{State Key Laboratory of Theoretical Physics, Institute of
  Theoretical  Physics, Chinese  Academy of  Sciences, P.O.  Box 2735,
  Beijing 100190, China}
%|--------------------------------------------------------------------|

% ----> DATE

\date{\today}

% ----> ABSTRACT

%|--------------------------------------------------------------------|
\begin{abstract}
  % == Problem ==
  % |--------------------------------------------------------------------|
  In the near future, gravitational wave detection is set to become an
  important observational  tool for  astrophysics. It will  provide us
  with  an  excellent  means  to distinguish  different  gravitational
  theories.   In effective  form, many  gravitational theories  can be
  cast into an \fR theory.  In this article, we study the dynamics and
  gravitational waveform of an  equal-mass binary black hole system in
  $f(R)$ theory.
  % ==                            Method                            ==
  % |--------------------------------------------------------------------|
  We  reduce   the  equations   of  motion  in   \fR  theory   to  the
  Einstein-Klein-Gordon  coupled  equations.   In  this  form,  it  is
  straightforward to modify  our existing numerical relativistic codes
  to simulate binary black hole  mergers in \fR theory.  We considered
  binary black holes surrounded by  a shell of scalar field.  We solve
  the  initial  data  numerically   using  the  \Olliptic  code.   The
  evolution  part is  calculated  using the  extended \AMSSNCKU  code.
  Both codes  were updated and tested  to solve the  problem of binary
  black holes in \fR theory.
  % == Results ==
  % |--------------------------------------------------------------------|
  Our  results show  that the  binary  black hole  dynamics in  $f(R)$
  theory is  more complex than in general  relativity.  In particular,
  the  trajectory and  merger  time are  strongly  affected.  Via  the
  gravitational wave,  it is possible to constrain  the quadratic part
  parameter  of \fR  theory in  the range  $|a_2|<10^{11}$m${}^2$.  In
  principle, a  gravitational wave detector can  distinguish between a
  merger of binary black hole  in \fR theory and the respective merger
  in   general  relativity.    Moreover,   it  is   possible  to   use
  gravitational wave detection to distinguish between \fR theory and a
  non self-interacting scalar field model in general relativity.
  % |--------------------------------------------------------------------|
\end{abstract}
% |--------------------------------------------------------------------|

% ----> PACS


% ----> MAKETITLE <----

\maketitle

%\tableofcontents

% -------------------------------------
%           ARTICLE
% -------------------------------------

% ----> INTRODUCTION

\section{introduction}
\label{sec:introduction}

%|--------------------------------------------------------------------|
Einstein's general  relativity (GR)  is currently the  most successful
gravitational   theory.   It   has  excellent   agreement   with  many
experiments (see e.g.~\cite{Wil06,Sta03,Ash03}).  However, most of the
tests involve  weak gravitational fields.   On the other  hand, recent
cosmological observations  require ad-hoc  explanations to fit  in the
framework of  GR theory, for example  the dark energy  and dark matter
problems \cite{Hut10,CopSamTsu06,LiLiWan11}.  In  order to solve these
difficulties,  some  alternative   gravitational  theories  have  been
proposed \cite{JaiKho10,SilTro09}.
%|--------------------------------------------------------------------|

%|--------------------------------------------------------------------|
In effective  form, many gravitational  theories can be caste  into an
\fR                                                              theory
\cite{FelTsu10,SotFar10,HuaLiMa10,ZhaMa11,ZhaMa11a}. Additionally, \fR
theory  has  a  relatively  simple  form.  Therefore,  it  is  a  good
alternative gravitational  model.  In  this work, we  characterize the
gravitational waveform of binary black hole mergers in \fR theory.
%|--------------------------------------------------------------------|

%|--------------------------------------------------------------------|
In  the  near future,  gravitational  wave  detection  will become  an
observational            method            for            astrophysics
\cite{AbrAltAre92,BraFabVir90,Mag00,GonXuBai11}.    The  gravitational
wave experiments can be excellent tools for testing GR in strong field
regime.   Moreover,  it  will  be possible  to  distinguish  different
gravitational theories.
%|--------------------------------------------------------------------|
Quantitatively, future experimental data  can be used to constrain \fR
parameters,  and to  confirm  or to  reject alternative  gravitational
theories.  With  this in  mind, we analyze  the waveforms in  order to
quantify the differences.  According to our results, it is possible to
distinguish quadratic  models of \fR  and GR with  future experimental
data.
%|--------------------------------------------------------------------|

%|--------------------------------------------------------------------|
The quadratic form of \fR is  given by $f(R)=R+a_2 R^2$. The main free
parameter is the coefficient of  the quadratic part $a_2$. In the case
$a_2=0$, \fR theory reduces to GR. In linearized \fR it is possible to
shows that Mercury's orbit sets the value of $\vert a_2 \vert \leq 1.2
\times 10^{18}  \, \mathrm{m}^2$ \cite{BerGai11}.  On  the other hand,
E\"ot-Wash experiments restrict  the value of $\vert a_2  \vert \leq 2
\times   10^{-9}\,\mathrm{m}^2$  \cite{HoyKapHec04,KapCooAde07}.   The
Laser Interferometer Space Antenna  (LISA) may distinguish $|a_2| \geq
10^{17}\,\mathrm{m}^2$.  Binary black holes  in the mass range $30-300
\, M_{\mathrm{sun}}$ are expected to  merge at frequencies in the most
sensitive  region  of  the  Laser  Interferometer  Gravitational  Wave
Observatory (LIGO)  frequency band \cite{VaiHinSho09}.   Therefore, we
focused our attention  on an equal-mass binary black  hole system with
total mass  $M=m_1+m_2=100\,M_{\mathrm{sun}}$.  We find  that the LIGO
detection can distinguish $|a_2| \geq 10^{11}\,\mathrm{m}^2$.
%|--------------------------------------------------------------------|

%|--------------------------------------------------------------------|
The paper  is organized as follows:  in Sec.~\ref{sec:math-backgr}, we
summarize  the  equations  of  \fR  theory.  This  is  followed  by  a
description of the initial  data setup in Sec.~\ref{initial-data}.  In
Sec.~\ref{sec:numerical-method-2},    we   describe    the   numerical
techniques   used   to   solve    the   equations   of   motion.    In
Sec.~\ref{sec:results-2}, we  give some motivation  and background for
the  configuration used  in this  work.  The  evolution  of equal-mass
binary      black      hole      system      is      presented      in
Sec.~\ref{sec:results-3}. Conclusions and discussions are presented in
Sec.~\ref{sec:discussion}.
%|--------------------------------------------------------------------|

\subsection{Notation and units}
\label{sec:notation-units}

%|--------------------------------------------------------------------|
We  employ  the following  notation:  Space-time  indices take  values
between 0 and  3, with 0 representing the  time coordinate.  The first
Latin  indices  $(a,  b,  c,  \dots,  h)$  refer  to  four-dimensional
space-time and take  values between 0 and 3,  while Latin indices $(i,
j, k, l, \dots)$ refer to three-dimensional space and take values from
1 to 3. The metric  signature is $(-1,1,1,1)$.  Some references (e.g.,
\cite{BerGai11}),   use  a   metric  signature   $(1,-1,-1,-1)$.   The
difference is a change of sign  of the scalar curvature $R$ as well as
$f(R)$.  We  use Einstein's  summation convention.  The  symbol $a:=b$
means  that  $a$  is  defined  as  being b.   A  dot  over  a  symbol,
$\dot{\vec{x}}$,  means   the  total  time   derivative,  and  partial
differentiation  with respect  to  $x^i$ is  denoted by  $\partial_i$.
Differentiation with respect to the Ricci scalar $R$ is denoted with a
prime, for example $f':=\frac{df(R)}{d R}$.
%|--------------------------------------------------------------------|

%|--------------------------------------------------------------------|
In order to  simplify the calculations, we use  geometric units, where
the  speed  of  light  $c$  and the  gravitational  constant  $G$  are
normalized to 1.  A variable  in bold font, i.e. $\mathbf{x}$, denotes
physical quantities  in international system  units. Particularly, the
values  of   $a_2  \approx   1\,  \mathrm{M}^2$  in   geometric  units
corresponds to $\mathbf{a_2} \approx 10^{11} \mathrm{m}^2$ for typical
gravitational wave sources of binary black hole for LIGO.
%|--------------------------------------------------------------------|

%|--------------------------------------------------------------------|
We  use the  following abbreviations:  Einstein's  general relativity
(GR), Laser Interferometer  Space Antenna (LISA), Laser Interferometer
Gravitational  Wave Observatory  (LIGO),  Einstein-Klein-Gordon (EKG),
Baumgarte-Shapiro-Shibata-Nakamura (BSSN), Arnowitt-Deser-Misner (ADM)
and binary black hole (BBH).
%|--------------------------------------------------------------------|

% ----> DYNAMICAL EQUATION OF f(R) THEORY FOR NUMERICAL RELATIVITY

%\section{dynamical equations of \fR theory for numerical relativity}
\section{Mathematical background}
\label{sec:math-backgr}

%|--------------------------------------------------------------------|
In  vacuum  spacetimes, \fR  theory  generalizes the  Hilbert-Einstein
action to
%|--------------------------------------------------------------------|
\begin{equation}
S=\int\frac{d^4x}{16\pi}\sqrt{-g}f(R),\label{eq:1}
\end{equation}
%|--------------------------------------------------------------------|
where GR is recovered by setting $f(R)=R$. From this action, we obtain
the Euler-Lagrange equations of motion
%|--------------------------------------------------------------------|
\begin{equation}
f^{\prime}R_{a b }-\frac{1}{2}fg_{a b }-[\nabla_{a }\nabla_{b }-g_{a
b }\Box]f^{\prime}=0.\label{eq:2}
\end{equation}
%|--------------------------------------------------------------------|
Using the definition of Einstein tensor $G_{ab}:=R_{ab}-g_{ab}R/2$, we
obtain  after   subtracting  a   Ricci  tensor  term   $Rg_{ab}/2$  in
\eqref{eq:2}, and rearranging terms,
%|--------------------------------------------------------------------|
\begin{equation}
G_{ab}=\frac{1}{f^{\prime}}\left[\nabla_a\nabla_b
f^{\prime}-g_{ab}\Box
f^{\prime}-\frac{1}{2}g_{ab}\left(Rf^{\prime}-f\right)\right].\label{eq:3}
\end{equation}
%|--------------------------------------------------------------------|
On   the  other   hand,  considering   the   conformal  transformation
$\tilde{g}_{ab}=e^{2\omega}g_{ab}$, the Ricci tensor transforms into
%|--------------------------------------------------------------------|
\begin{align}
&\tilde{R}_{ab}=R_{ab}-\nonumber\\
&2\nabla_a\nabla_b\omega-g_{ab}\Box\omega+2\nabla_a\omega\nabla_b\omega
-2g_{ab}g^{de}\nabla_d\omega\nabla_e\omega.\label{eq:4}
\end{align}
%|--------------------------------------------------------------------|
The corresponding Ricci scalar transforms as
%|--------------------------------------------------------------------|
\begin{equation}
\tilde{R}=e^{-2\omega}\left(R-6\Box\omega-6g^{de}\nabla_d\omega\nabla_e\omega\right).\label{eq:5}
\end{equation}
%|--------------------------------------------------------------------|
Therefore, the Einstein tensor transformation is given by
%|--------------------------------------------------------------------|
\begin{align}
&\tilde{G}_{ab}=G_{ab}-\nonumber\\
&2\nabla_a\nabla_b\omega+2g_{ab}\Box\omega+2\nabla_a\omega\nabla_b\omega+
g_{ab}g^{de}\nabla_d\omega\nabla_e\omega.\label{eq:6}
\end{align}

%|--------------------------------------------------------------------|
Defining $\omega:=\frac{1}{2}\ln\lambda$, we have
%|--------------------------------------------------------------------|
\begin{align}
\nabla_a\omega=&\frac{1}{2\lambda}\nabla_a\lambda,\label{eq:7}\\
\nabla_a\nabla_b\omega=&-\frac{1}{2\lambda^2}\nabla_a\lambda\nabla_b\lambda+\frac{1}{2\lambda}\nabla_a\nabla_b\lambda.\label{eq:8}
\end{align}
%|--------------------------------------------------------------------|
The substitution  of \eqref{eq:7} and  \eqref{eq:8} in \eqref{eq:6}
implies
%|--------------------------------------------------------------------|
\begin{align}
\tilde{G}_{ab}=&G_{ab}+\frac{3}{2\lambda^2}\nabla_a\lambda\nabla_b\lambda-\frac{3}{4\lambda^2}g_{ab}g^{de}\nabla_d\lambda\nabla_e\lambda\nonumber\\
&-\frac{1}{\lambda}\left(\nabla_a\nabla_b\lambda-g_{ab}\Box\lambda\right).\label{eq:9}
\end{align}
%|--------------------------------------------------------------------|
Substituting    $\lambda:=f^{\prime}$   in   \eqref{eq:3}    and   the
result in \eqref{eq:9}, we get
%|--------------------------------------------------------------------|
\begin{equation}
\tilde{G}_{ab}=\frac{3}{2\lambda^2}\nabla_a\lambda\nabla_b\lambda-\frac{3}{4\lambda^2}g_{ab}g^{de}\nabla_d\lambda\nabla_e\lambda
-\frac{\left(R\lambda-f\right)}{2\lambda}g_{ab}.\label{eq:10}
\end{equation}
%|--------------------------------------------------------------------|
Since  the conformal transformation  satisfies $\tilde{g}_{ab}=\lambda
g_{ab}$, \eqref{eq:10} takes the form
%|--------------------------------------------------------------------|
\begin{equation}
\tilde{G}_{ab}=\frac{3}{2\lambda^2}\tilde{\nabla}_a\lambda\tilde{\nabla}_b\lambda-\frac{3}{4\lambda^2}\tilde{g}_{ab}\tilde{g}^{de}
\tilde{\nabla}_d\lambda\tilde{\nabla}_e\lambda
-\frac{\left(R\lambda-f\right)}{2\lambda^2}\tilde{g}_{ab}.\label{eq:11}
\end{equation}
%|--------------------------------------------------------------------|
Defining $\phi:=\sqrt{\frac{3}{16\pi}}\ln\lambda$, we get
%|--------------------------------------------------------------------|
\begin{equation}
  \tilde{G}_{ab}=8\pi\left[\tilde{\nabla}_a\phi\tilde{\nabla}_b\phi-\tilde{g}_{ab}\left(\frac{1}{2}\tilde{g}^{de}
      \tilde{\nabla}_d\phi\tilde{\nabla}_e\phi
      +V\right)\right].\label{eq:12}
\end{equation}
%|--------------------------------------------------------------------|
where           
\begin{equation}
 V:=\frac{R \e^{4\sqrt{\pi/3}\phi}-f}{16\pi \e^{8\sqrt{\pi/3}\phi}}.
\end{equation}
  The  right hand side  of \eqref{eq:12}  has the  form of  the stress
  energy tensor of a scalar field (see e.g.~\cite{Wal84,Car03})
%|--------------------------------------------------------------------|
\begin{equation}
\tilde{T}_{ab}:=\tilde{\nabla}_a\phi\tilde{\nabla}_b\phi-\tilde{g}_{ab}
\left(\frac{1}{2}\tilde{\nabla}_c\phi\tilde{\nabla}^c\phi+V\right).\label{eq:13}
\end{equation}
%|--------------------------------------------------------------------|
Therefore,  in  vacuum,  the  \fR  theory  equations  of  motion are
equivalent to GR equations coupled to a real scalar field
%|--------------------------------------------------------------------|
\begin{equation}
\phi = \frac{\sqrt{3}}{4\sqrt{\pi}}\ln f'.\label{eq:14}
\end{equation}
%|--------------------------------------------------------------------|

%|--------------------------------------------------------------------|
The equation  of motion of the scalar  field is given by  the trace
of \eqref{eq:2} with $g^{ab}$
%|--------------------------------------------------------------------|
\begin{equation}
\tilde{\square}f'=2\tilde{\nabla}^a\omega\tilde{\nabla}_af'-\frac{2f-f'R}{3},\label{eq:15}
\end{equation}
%|--------------------------------------------------------------------|
where   we  have   employed  the   conformal   metric  transformation.
Substituting the definition of $\phi$ we get
%|--------------------------------------------------------------------|
\begin{align}
\tilde{\square}\phi&=\frac{2f-f'R}{4\sqrt{3\pi}f'^2}\nonumber\\
&=\frac{2f-Rf'}{16\pi f'^3}f''\frac{dR}{d\phi}\nonumber\\
&=\frac{dV}{d\phi}.\label{eq:16}
\end{align}
%|--------------------------------------------------------------------|
The  result is  the dynamical  equation of  a real  scalar  field with
potential $V$. Therefore,  the equations of motion for  \fR theory are
equivalent to Eqs.~\eqref{eq:12} and \eqref{eq:16}, which form the EKG
system of equations.   Notice that the scalar field  is introduced for
numerical simulation convenience. Moreover, it is related to the Ricci
scalar. Therefore, it does not represent a physical freedom.
%|--------------------------------------------------------------------|

%|--------------------------------------------------------------------|
The equations  of motion derived with the  metric $\tilde{g}_{ab}$ are
commonly  referred  to  be   in  the  Einstein  frame.   For  physical
interpretation, we  need to transform  them using the  physical metric
$g_{ab}   =   e^{-4\sqrt{\frac{\pi}{3}}\phi}   \tilde{g}_{ab}$.    The
equations in that form are referred  to be in the Jordan frame. We use
Newman-Penrose  scalar  $\Psi_4$  to analyze  gravitational  waveform.
Therefore,          it          is         calculated          through
$\tilde{\Psi}_4=e^{-4\sqrt{\frac{\pi}{3}}\phi}\Psi_4$.  Since the Weyl
tensor  is conformal  invariant, the  pre-factor comes  from  a tetrad
transformation.
%|--------------------------------------------------------------------|

%|--------------------------------------------------------------------|
We use  3+1 formalism to  solve \eqref{eq:12} and  \eqref{eq:16}.  For
Einstein equations  \eqref{eq:12} we adopt the BSSN  formulation as in
our  previous  work  \cite{CaoYoYu08}.   The  scalar  field  equations
\eqref{eq:16}  can be decomposed  using the  3+1 formalism  as follows
(see  e.g., for  detail about  the 3+1  formalism \cite{Alc08,Gou12}):
First    it   is    useful   to    define   an    auxiliary   variable
$\varphi:=\DLie_{\mathbf{n}} \phi$, where $\DLie_{\mathbf{n}}$ denotes
the Lie  derivative along the  normal to the  hypersurface $\Sigma_t$.
Expressing the Lie derivative in  terms of the lapse function $\alpha$
and the shift vector $\beta^i$, the evolution of $\phi$ is given by
%|--------------------------------------------------------------------|
\begin{equation}
\partial_t\phi =\alpha\varphi+\beta^i\partial_i\phi.\label{eq:17}
\end{equation}
%|--------------------------------------------------------------------|
On  the  other  hand, the  evolution  of  $\varphi$  is given  by
the substitution of $\DLie_{\mathbf{n}}\phi$ in \eqref{eq:16}
%|--------------------------------------------------------------------|
\begin{align}
\partial_t \varphi &=\alpha\chi\left(\bar{\gamma}^{ij}\partial_i \partial_j\phi-(\bar{\Gamma}^i+\frac{\bar{\gamma}^{ij}\partial_j\chi}{2\chi})\partial_i\phi \right)
+\chi\bar{\gamma}^{ij}\partial_i\alpha \partial_j\phi\nonumber\\
&+\alpha\varphi
K-\alpha\frac{dV}{d\phi}+\beta^i\partial_i\varphi,\label{eq:18}
\end{align}
%|--------------------------------------------------------------------|
where    we   used   the    BSSN   metric    conformal   transformation
$\bar{\gamma}_{ij}=\chi \gamma_{ij}$ and the relationships
%|--------------------------------------------------------------------|
\begin{align}
K&=-\frac{\gamma^{ij}}{2\alpha}\frac{\partial \gamma_{ij}}{\partial t},\\
\Gamma^i&= -\frac{1}{\sqrt{\gamma}}\partial_j\left(\sqrt{\gamma}\gamma^{ij}\right),\label{eq:19}
\end{align}
%|--------------------------------------------------------------------|
with  $K$   the  trace  of  the  extrinsic   curvature,  $\gamma$  the
determinant of the 3-metric  and $\Gamma^i$ the contracted Christoffel
symbol.   The quantities  with an  upper  bar are  represented in  the
conformal metric of BSSN form.
%|--------------------------------------------------------------------|

%|--------------------------------------------------------------------|
The matter densities are given by
%|--------------------------------------------------------------------|
\begin{align}
E &:= n_{a}n_{b}T^{a b}\nonumber\\
&=\frac{1}{2}D_i\phi D^i\phi+\frac{1}{2}\varphi^2+V,\label{eq:20}\\
p_i &:= -\gamma_{i a}n_{b}T^{a b}\nonumber\\
&=-\varphi D_i\phi,\label{eq:21}\\
S_{ij} &:= \gamma_{i a}\gamma_{j b}T^{ a b}\nonumber\\
&=D_i\phi
D_j\phi-\gamma_{ij}\left(\frac{1}{2}D_k\phi
D^k\phi-\frac{1}{2}\varphi^2+V\right).\label{eq:22}
\end{align}
%|--------------------------------------------------------------------|
For $f$, we consider a  quadratic form $f(R)=R+a_2 R^2$, which results
in the potential
%|--------------------------------------------------------------------|
\begin{equation}
V=\frac{1}{32\pi a_2}(1-e^{4\sqrt{\pi/3}\phi})^2e^{-8\sqrt{\pi/3}\phi}.
\end{equation}
%|--------------------------------------------------------------------|
This potential is analytic around $\phi=0$ and it can be expanded as
%|--------------------------------------------------------------------|
\begin{equation}
V=\frac{1}{6a_2}{\phi}^{2}-\frac{2}{3a_2}\sqrt{\frac{\pi}{3}}{\phi}^{3}+{\frac
{14\pi}{27a_2}}{\phi}^{
4}-\frac{8\pi}{9a_2}\sqrt{\frac{\pi}{3}}{\phi}^{5}+O \left(
{\phi}^{6} \right).
\end{equation}
%|--------------------------------------------------------------------|
The coefficient of $\phi^2$ is related to the mass of the scalar field
($m=1/\sqrt{6a_2}$) and  the other terms  imply that the  scalar field
has nonlinear  self-interaction.  With the  signature convention taken
in  this  work, only  the  positive  values  of $a_2$  are  physically
meaningful. Therefore, we demand that $a_2 \geq 0$.
%|--------------------------------------------------------------------|

\subsection{Formalism for numerical calculation of \fR dynamics}

%|--------------------------------------------------------------------|
The dynamical equations for \fR theory can be written as (\ref{eq:2}),
or equivalently  as (\ref{eq:12}).  There  is a key component  in BSSN
formalism where  $\bar{\Gamma}^i$ are  consider to be  new independent
functions. Similar  to this,  we promote $\phi$  to a  new independent
function.   Then the  evolution equation  of $\phi$  is  determined by
(\ref{eq:16}).   On   the  other   hand,  the  definition   of  $\phi$
(\ref{eq:14}  is  a  constraint  equation.  For  later  reference,  we
summarize the  equations for numerical calculation of  \fR dynamics as
follows
\begin{align}
\tilde{G}_{ab}&=8\pi \tilde{T}_{ab},\label{eq:23}\\
\tilde{\square}\phi&=\frac{dV}{d\phi}.\label{eq:24}
\end{align}
The constraint equation is
\begin{align}
\ln f'=\frac{4\sqrt{\pi}}{\sqrt{3}}\phi.\label{eq:25}
\end{align}
%|--------------------------------------------------------------------|

%|--------------------------------------------------------------------|
It  is  interesting  to  note  that the  original  dynamical  equation
(\ref{eq:2})  for \fR theory  includes 4th  order derivative  terms of
metric.  This  is because  $f$ depends on  $R$, which  contains second
derivative  terms  of the  metric,  and  \eqref{eq:2} contains  second
derivative terms of $f$.  After performing a conformal transformation,
we obtain  the dynamical  equation \eqref{eq:12}.  If  we look  at the
conformal  metric $\tilde{g}_{ab}$  instead of  $g_{ab}$  as dynamical
variables,  (\ref{eq:12}) involves  3rd order  derivatives  which come
from  the derivative of  $\phi$. This  is because  $\phi$ itself  is a
function  of  $R$  which   contains  second  derivative  of  conformal
metric. In  (\ref{eq:23}) and (\ref{eq:24}), we replace  the 3rd order
derivative  terms by  promoting the  auxiliary variable  $\phi$  as an
independent  variable. This treatment  introduces an  extra constraint
equation  (\ref{eq:25}) which  is similar  to  the role  of the  Gamma
constraint equations  in BSSN numerical scheme.   With this treatment,
equations (\ref{eq:23}) and (\ref{eq:24}) contain at most second order
derivative terms.
%|--------------------------------------------------------------------|

%|--------------------------------------------------------------------|
The system of equations (\ref{eq:23}) and (\ref{eq:24}) takes the form
of coupled Einstein-Klein-Gordon  equations.  For Einstein equation we
use  the  BSSN  formulation.    We  monitor  the  constraint  equation
(\ref{eq:25}) to check the consistency of our numerical solutions.
%|--------------------------------------------------------------------|

% ----> INITIAL DATA

\section{Initial Data}
\label{initial-data}

%|--------------------------------------------------------------------|
Under a 3+1 decomposition, the constraint equations read as follows:
%|--------------------------------------------------------------------|
\begin{align}
  D_j K^j{}_i - D_i K  &= 8\pi p_i,\label{eq:26} \\
  R + K^2 - K_{ij} K^{ij} &= 16 \pi E, \label{eq:27}
\end{align}
%|--------------------------------------------------------------------|
where $R$  is the Ricci  scalar, $K_{ij}$ is the  extrinsic curvature,
$K$ the trace of  the extrinsic curvature, $\gamma_{ij}$ the 3-metric,
and $D_j$ the covariant derivative associated with $\gamma_{ij}$.  $E$
and $p_i$  are the  energy and momentum  densities given  in equations
\eqref{eq:20} and \eqref{eq:21}.
%|--------------------------------------------------------------------|

% ----> PUNCTURES METHOD

\subsection{Puncture method}
\label{sec:punctures}

%|--------------------------------------------------------------------|
The   constraints   can   be   solved   with   the   puncture   method
\cite{BraBru97}.    Following   the   conformal   transverse-traceless
decomposition  approach, we  make  the following  assumptions for  the
metric and the extrinsic curvature:
%|--------------------------------------------------------------------|
\begin{align}
  &\gamma_{ij} =  \psi_0^4 \hat{\gamma}_{ij},& \label{eq:28} \\
  &K_{ij} = \psi_0^{-2} \hat{A}_{ij}
  + \frac{1}{3} K \hat{\gamma}_{ij},&\label{eq:29}
\end{align}
%|--------------------------------------------------------------------|
where $\hat{A}_{ij}$ is trace free and $\psi_0$ is a conformal factor.
We     chose     a     conformally     flat     background     metric,
$\hat{\gamma}_{ij}=\delta_{ij}$,   and  a  maximal   slice  condition,
$K=0$.     The     last     choice    decouples     the     constraint
equations~\eqref{eq:26}-\eqref{eq:27} to take the form
%|--------------------------------------------------------------------|
\begin{align}
  &  \partial_j \hat{A}^{ij} = 0, & \label{eq:30} \\
  & \vartriangle \psi_0 + \frac{1}{8} \hat{A}^{ij} \hat{A}_{ij}
  \psi_0^{-7} =-\psi_0 \delta^{ij}\partial_i \phi \partial_j \phi - 2 \pi \psi_0^5 V, & \label{eq:31}
\end{align}
%|--------------------------------------------------------------------|
where  $\vartriangle$  is   the  Laplacian  operator  associated
with Euclidian     metric.      Notice     that     we      have
chosen $\varphi\equiv\DLie_{\mathbf{n}}\phi   =  0$  initially. This
is consistent to the quasi-equilibrium  picture. So $p_i=0$ which
results in \eqref{eq:30}.
%|--------------------------------------------------------------------|

%|--------------------------------------------------------------------|
In  a   Cartesian  coordinate  system  $(x^i)=(x,y,z)$,   there  is  a
non-trivial solution of \eqref{eq:30} which is valid for any number of
black holes  \cite{BowYor80} (here the index  $n$ is a  label for each
puncture):
%|--------------------------------------------------------------------|
\begin{align}
  \hat{A}^{ij} &=& \sum_n \left[ \frac{3}{2 r^3_n} \left[ x_n^i P_{n}^{j}
      + x_n^j P_{n}^{i} - \left(  \delta^{ij}
      - \frac{ x_n^i x_n^j}{r_n^2} \right) P^n_k x_n^k \right]
    \right.\nonumber \\&& \left.
    + \frac{3}{r_n^5} \left(  \epsilon^{ik}_{\;\;l} S^n_k x_n^l  x_n^j +
    \epsilon^{jk}_{\;\;l} S^n_k x_n^l x_n^i \right) \right],\label{eq:32}
\end{align}
%|--------------------------------------------------------------------|
where                      $r_n:=\sqrt{(x-x_n)^2+(y-y_n)^2+(z-z_n)^2}$,
$\epsilon^{ik}_{\;\;l}$ is the  Levi-Civita tensor associated with the
flat  metric, and  $P_n$  and $S_n$  are  the ADM  linear and  angular
momentum of $n$th black hole, respectively.
%|--------------------------------------------------------------------|

%|--------------------------------------------------------------------|
The Hamiltonian constraint  \eqref{eq:31} becomes an elliptic equation
for the conformal  factor $\psi_0$. The solution splits as  a sum of a
singular term and a finite correction $u$ \cite{BraBru97},
%|--------------------------------------------------------------------|
\begin{equation}
  \psi_0 = 1 + \sum_n \frac{m_n}{2r_n} + u, \label{eq:33}
\end{equation}
%|--------------------------------------------------------------------|
with $u \rightarrow 0$ as  $r_n \rightarrow \infty$.  The function $u$
is determined by an elliptic equation on $\R^3$, which is derived from
\eqref{eq:31}  by  inserting  \eqref{eq:33}, and  $u$  is  $C^\infty$
everywhere except at the punctures,  where it is $C^2$.  The parameter
$m_n$ is called the bare mass of the $n$th puncture.
%|--------------------------------------------------------------------|

% ----> NUMERICAL METHOD

\subsection{Numerical Method}
\label{sec:numerical-method-1}

%|--------------------------------------------------------------------|
The Hamiltonian  constraint \eqref{eq:31} is  solved numerically using
the  \Olliptic code  (\cite{GalBruCao10a}).  \Olliptic  is  a parallel
computational code which solves three dimensional systems of nonlinear
elliptic  equations  with  a  2nd,  4th, 6th,  and  8th  order  finite
difference                       multigrid                      method
\cite{Bra77,BaiBra87,BraLan88,HawMat03,ChoUnr86b}.     The    elliptic
solver uses vertex-centered stencils and box-based mesh refinement.
%|--------------------------------------------------------------------|

%|--------------------------------------------------------------------|
The numerical domain is represented by a hierarchy of nested Cartesian
grids. The hierarchy consists of $L+G$ levels of refinement indexed by
$l = 0, \ldots ,L + G - 1$. A refinement level consists of one or more
Cartesian  grids with  constant grid-spacing  $h_l$ on  level  $l$.  A
refinement factor of two is used such that $h_l = h_G/2^{|l-G|}$.  The
grids are properly nested in that the coordinate extent of any grid at
level $l > G$ is completely  covered by the grids at level $l-1$.  The
level $l=G$  is the  ``external box'' where  the physical  boundary is
defined.  We  used grids with  $l<G$ to implement the  multigrid method
beyond level $l=G$.
%|--------------------------------------------------------------------|

%|--------------------------------------------------------------------|
For  the  outer  boundary,  we  required  an  inverse  power  fall-off
condition,
%|--------------------------------------------------------------------|
\begin{equation}
  u(r)=A+\frac{B}{r^q}, \quad \mathrm{for} \quad r \gg 1, \; q > 0, \label{eq:34}
\end{equation}
%|--------------------------------------------------------------------|
where the factor  $B$ is unknown. It is possible  to get an equivalent
condition which does not contain  $B$ by calculating the derivative of
\eqref{eq:34} with respect to  $r$, solving the equation for $B$
and making  a substitution in the  original equation. The  result is a
\textit{Robin} boundary condition:
%|--------------------------------------------------------------------|
\begin{equation}
  u(\vec{x}) +  \frac{r}{q} \frac{\partial u(\vec{x})}{\partial r} = A.\label{eq:35}
\end{equation}
%|--------------------------------------------------------------------|
For the initial data, we set $q=1$ and $A=0$.

% ----> RESULTS

\subsection{Results}
\label{sec:results-1}

\subsubsection{Test problem}
\label{sec:test-problem}

%|--------------------------------------------------------------------|
As a  test, we set the  mass parameter of  the black hole to  zero and
consider a  spherical symmetric field  $\phi$ and potential  $V$.  The
Hamiltonian  constraint  \eqref{eq:31}   reduces  to  a  second  order
ordinary differential equation
%|--------------------------------------------------------------------|
\begin{equation}
r \psi_0'' + 2 \psi_0' + \pi \psi_0 (\phi')^2+2\pi V(r) \psi^5_0 = 0, \label{eq:36}
\end{equation}
% |--------------------------------------------------------------------|
where the prime denotes differentiation with respect to $r$.  In order
to   obtain   a   high-resolution   reference  solution,   we   solve
\eqref{eq:36}  using   \textsc{Mathematica}  \cite{Wol08}.   A  useful
transformation for  the case $V=0$ is $\psi_1:=r  \psi_0$.  Under this
transformation, regularity at  the origin implies $\lim_{r \rightarrow
  0}\psi_1(r)=0$. The boundary  condition \eqref{eq:35} with $q=1$ and
$A=1$    reduces     to    $    \psi_1'(r_{\mathrm{max}})=1$,    where
$r_{\mathrm{max}}$ is the radius of our numerical domain.  The problem
then becomes
%|--------------------------------------------------------------------|
\begin{align}
\psi_1'' + \pi \psi_1 (\phi')^2+2\pi V(r)\frac{\psi^5_1}{r^4} &= 0,\\
\psi_1(0) &= 0, \\
\psi_1'(r_{\mathrm{max}})&=1.
\end{align}
%|--------------------------------------------------------------------|
For the case  $V \neq 0$, the term $r^{-4}$  produces a singularity at
the  origin.  We  cure  artificially the  singularity  by solving  the
equation with  a term  $(r^4+\epsilon)^{-1}$ instead of  $r^{-4}$. For
the test, the value of $\epsilon$ is set to $10^{-12}$.
%|--------------------------------------------------------------------|

%|--------------------------------------------------------------------|
We considered 2 cases
%|--------------------------------------------------------------------|
\begin{equation}
\begin{array}{ll}
\mathrm{Case \quad I:} &\phi(r) = \phi_0 \tanh [(r-r_0)/\sigma], \\
&  V(r) = 0. \\
\mathrm{Case \quad II:} &\phi(r) = \phi_0 \e^{ -(r-r_0)^2/\sigma}, \\
&  V(r) = \frac{1}{32 \pi a_2}\left( 1- \e^{4\sqrt{\pi/3}\phi} \right)^2\e^{-8\sqrt{\pi/3}\phi},
\end{array}\nonumber
\end{equation}
%|--------------------------------------------------------------------|
where  in both cases  $r_0=120 M$,  $\sigma=8 M$,  $\phi_0=1/40$.  For
case II, we set $a_2= 1$.  The numerical domain is a cubic box of size
4000 ($r_{\mathrm{max}}=2000$) and 11  refinements levels.  We use the
fourth  order  finite difference  stencil  since  it  provides a  good
convergence  property   at  the   boundary  for  large   domains  (see
\cite{GalBruCao10a} for  details). The convergence tests  consist of a
set of six solutions with grid points $N_i \in \{43, 51, 75, 105, 129,
149  \}$.  The comparison  with the  reference solution  was performed
along the  $Y$ axis using  a 6th order Lagrangian  interpolation.  For
each  resolution, the  difference $E_i  := \vert  u_i  - \bar{u}\vert$
gives an  estimation of  the error.  Here  $u_i$ denotes  the solution
produced with \textsc{Olliptic}, $i$ is an index which labels the grid
size,  $\bar{u}$ the reference  solution and  $\vert \cdot  \vert$ the
absolute  value   (computed  point  by  point).    The  functions  are
interpolated in  a domain with  grid size $\Delta  y = 1$.   The error
satisfies  $E_i \sim  C h_i^p$,  where $C$  is a  constant,  $h_i \sim
1/N_i$ is the  grid size and $p$ the order  of convergence.  Using the
$L_1$ norm  of the  error and performing  a linear regression  of $\ln
\vert  E_i \vert_{L_1}$  vs $\ln  \vert  h_i \vert$,  we estimate  the
convergence order $p$ and the constant $C$.
%|--------------------------------------------------------------------|

\begin{figure}[tbp]
  \centering
  \includegraphics[width=85mm]{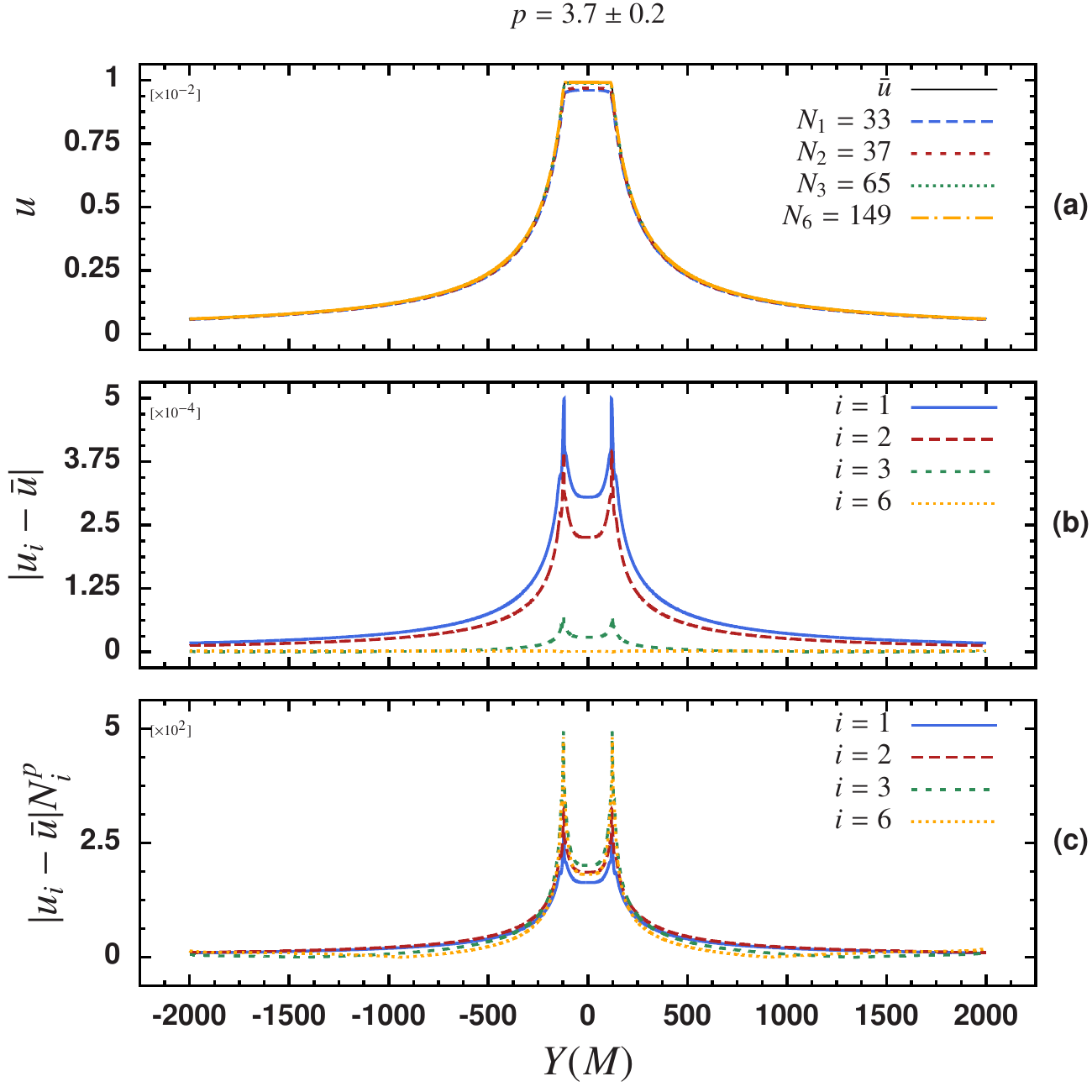}
  \caption{Initial  data convergence  test for case I.   The upper  panel
    \bfa, shows  the reference solution and 4  solutions computed with
    \Olliptic. The  middle panel  \bfb, presents the  estimated error.
    The lower panel \bfc, shows the scaled error for convergence order
    $p=3.7 \pm 0.2$.}\label{fig:1}
\end{figure}

%|--------------------------------------------------------------------|
Figure  \ref{fig:1} shows  the  result of  case  I.  There  is a  good
agreement between the several  resolutions and the reference solution.
The    plot    does    not    show   noticeable    differences    (see
Fig.~\ref{fig:1}-\bfa).  The solution  has convergence properties, and
the   estimated    error   diminishes   with    increased   resolution
(Fig.~\ref{fig:1}-\bfb).  The scaled error $E_i/h_i^p$ also shows good
convergence  with convergence order  $p=3.7 \pm  0.2$ given  by linear
regression (Fig.~\ref{fig:1}-\bfc).
%|--------------------------------------------------------------------|

%|--------------------------------------------------------------------|
%|--------------------------------------------------------------------|

\begin{figure}[tbp]
  \centering
  \includegraphics[width=85mm]{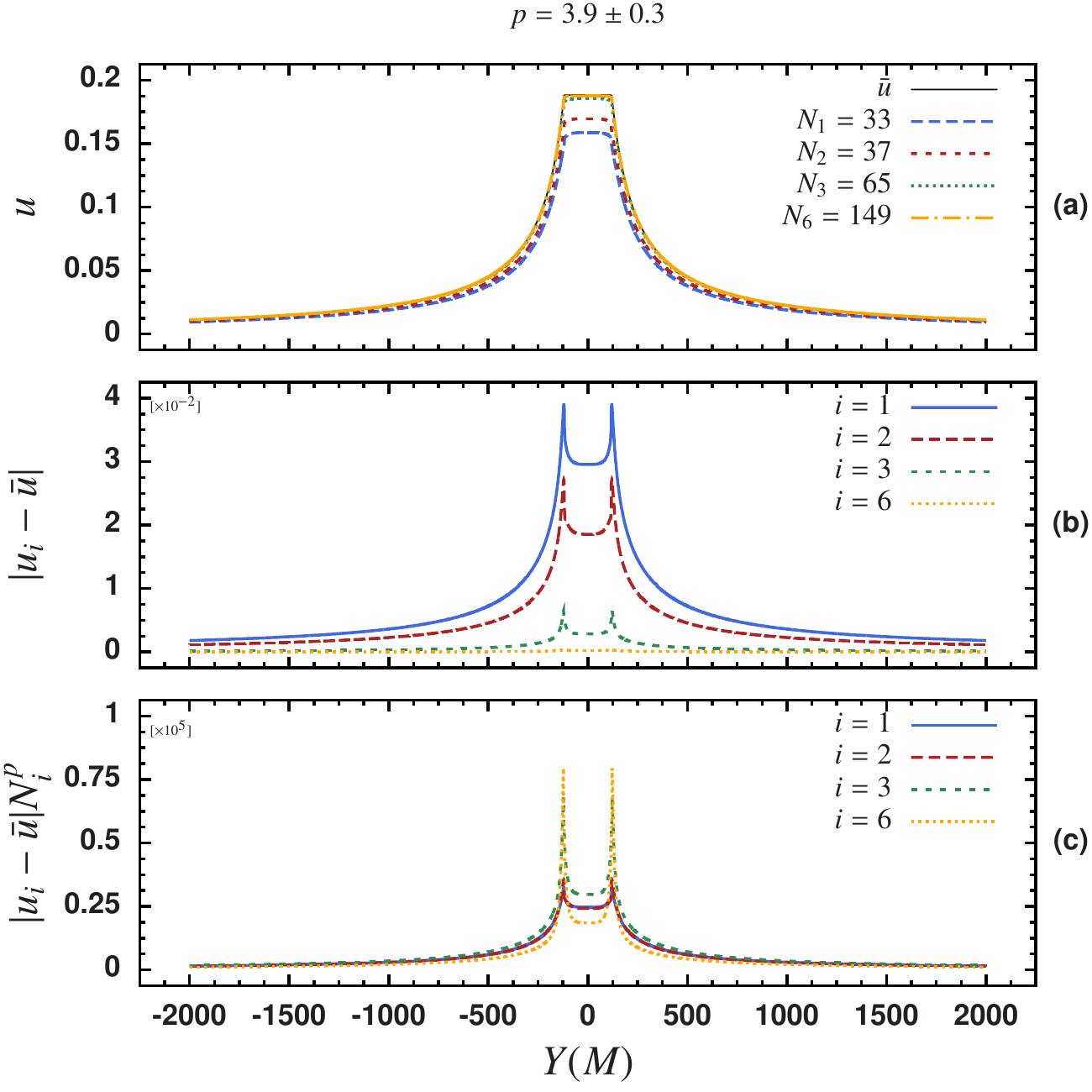}
  \caption{Initial  data convergence  test for case  II.  The  upper panel
    \bfa, shows  the reference solution and 4  solutions computed with
    \Olliptic. The  middle panel  \bfb, presents the  estimated error.
    The lower panel \bfc, shows the scaled error for convergence order
    $p=3.9 \pm 0.3$.}\label{fig:2}
\end{figure}

%|--------------------------------------------------------------------|
The  results for  case II  are presented  in  Figure~\ref{fig:2}.  The
solution is similar  to case I, an almost  constant solution between 0
and $r_0$ which joins a inverse power solution after $Y=r_0$. However,
the solution  of case II is  around 2 orders of  magnitude larger than
the  solution of case  I.  Contrary  to case  I, there  are noticeable
differences  between the reference  solution and  the lower-resolution
ones   (Fig.~\ref{fig:2}-\bfa).    The   solution  shows   convergence
properties and  the scaled  error shows convergence  consistently with
$p=3.9 \pm 0.3$ (Fig.~\ref{fig:2}-\bfb,\bfc).
%|--------------------------------------------------------------------|

%|--------------------------------------------------------------------|
%|--------------------------------------------------------------------|

\subsubsection{Initial data for evolution}
\label{sec:init-data-evol}

%|--------------------------------------------------------------------|
The  solution   of  \eqref{eq:31}   provides  initial  data   for  our
evolutions.   The initial  parameters of  the BBH  are:  puncture mass
parameter $m_1=m_2=0.487209$ (approximate apparent horizon mass equals
to 0.5),  initial position $(x,y,z)=(0,\pm5.5,0)$  and linear momentum
$(p_x,p_y,p_z)=(\mp0.0901099,\mp0.000703975,0)$.   The linear momentum
parameter is tuned for non-spinning quasi-circular orbits in GR.
%|--------------------------------------------------------------------|

%|--------------------------------------------------------------------|
\begin{table}[btp]
  \begin{ruledtabular}
    \caption{ADM  mass as  function of  $\phi_0$ for  $a_2 \rightarrow
      \infty$ (Fig.~\ref{fig:3}). The values are well represented by a
      quadratic function  $\MADM = A+B \phi_0^2$  with $A=0.99067$ and
      $B=40569 \pm 48$. }\label{tab:1}
    \begin{tabular}{l|cc||l|cc}
       \# & $\phi_0$  &   $\MADM$ & \# & $\phi_0$  &   $\MADM$  \\ \hline
      % |-------------------------------------------------------------|
       1 & 0       & 0.990669 & 7 & 0.004   & 1.632418\\
       2 & 0.0001  & 0.991069 & 8 & 0.005   & 1.994890\\
       3 & 0.0005  & 1.000670 & 9 & 0.006   & 2.439376\\
       4 & 0.001   & 1.030680 & 10 & 0.007   & 2.966764\\
       5 & 0.002   & 1.150790 & 11 & 0.008   & 3.578118\\
       6 & 0.003   & 1.351237 & 12 & 0.009   & 4.274675\\
      %|--------------------------------------------------------------|
    \end{tabular}
  \end{ruledtabular}
\end{table}
%|--------------------------------------------------------------------|

%|--------------------------------------------------------------------|
\begin{table}[btp]
  \begin{ruledtabular}
    \caption{ADM  mass as  function of  $a_2$  (Fig.~\ref{fig:4}).  The
      value of the maximum (\#  8) is estimated using the minimization
      of \eqref{eq:38}. The parameter  $\phi_0$ of the scalar field is
      $0.001642$.}\label{tab:2}
    \begin{tabular}{l|rc||l|rc}
       \# & $a_2 $  &   $\MADM$  & \# & $a_2$  &
      $\MADM$ \\ \hline
      % |-------------------------------------------------------------|
      1 & 0       & 9.906691 & 9 & 4       & 1.153111\\
      2 & 0.2     & 9.930327 & 10 & 6       & 1.140796\\
      3 & 0.4     & 1.006901 & 11 & 8       & 1.132395\\
      4 & 0.6     & 1.033066 & 12 & 10      & 1.126691\\
      5 & 0.8     & 1.063929 & 13 & 20      & 1.113947\\
      6 & 1       & 1.092333 & 14 & 40      & 1.106991\\
      7 & 2       & 1.155675 & 15 & 60      & 1.104598\\
      8 & 2.64791 & 1.160240 & 16 & 80      & 1.103388\\
      %|--------------------------------------------------------------|
    \end{tabular}
  \end{ruledtabular}
\end{table}
%|--------------------------------------------------------------------|

%|--------------------------------------------------------------------|
For the scalar field part, we consider that the BBH is surrounded by a
shell of scalar field with initial profile
%|--------------------------------------------------------------------|
\begin{equation}
\phi(r)=\frac{a_2^2}{a_2^2+1}\phi_0e^{-(r-r_0)^2/\sigma},\label{eq:37}
\end{equation}
%|--------------------------------------------------------------------|
with  $r_0=120M$,  $\sigma=8M$ and  several  values  of $\phi_0$  (see
below).  When  $a_2$ goes  to zero,  both $\phi$ and  $V$ go  to zero.
Therefore,  standard general  relativity is  recovered.  On  the other
hand, when $a_2 \rightarrow \infty$, the amplitude of the scalar field
tends to $\phi_0$ while the  potential vanishes. Our model provides an
unified  scheme  to  investigate  standard  GR  ($a_2=0$),  usual  \fR
($0<a_2<\infty$)  and the  free  EKG system  in  GR ($a_2  \rightarrow
\infty$).
% |--------------------------------------------------------------------|

%|--------------------------------------------------------------------|
From the solution  of the conformal factor it  is possible to estimate
the ADM mass through
%|--------------------------------------------------------------------|
\begin{equation}
\MADM \vert_{r=r_0} =-\frac{1}{2\pi}\oint_S \partial_j \psi\; \rd S^j,
\end{equation}
%|--------------------------------------------------------------------|
where the  integration is  performed in a  sphere $S$ of  radius $r_0$
(formally the ADM  mass is computed taking the  limit $r_0 \rightarrow
\infty$).  In  our calculations  $r_0=1537.5$ and the  integrations is
done numerically using 6th  order Lagrange interpolation in the sphere
and 6th order Boole's quadrature \cite{PreFlaTeu92a,KarKir03}.
%|--------------------------------------------------------------------|

%|--------------------------------------------------------------------|
\begin{figure}[tbp]
  \centering
  \includegraphics[width=85mm]{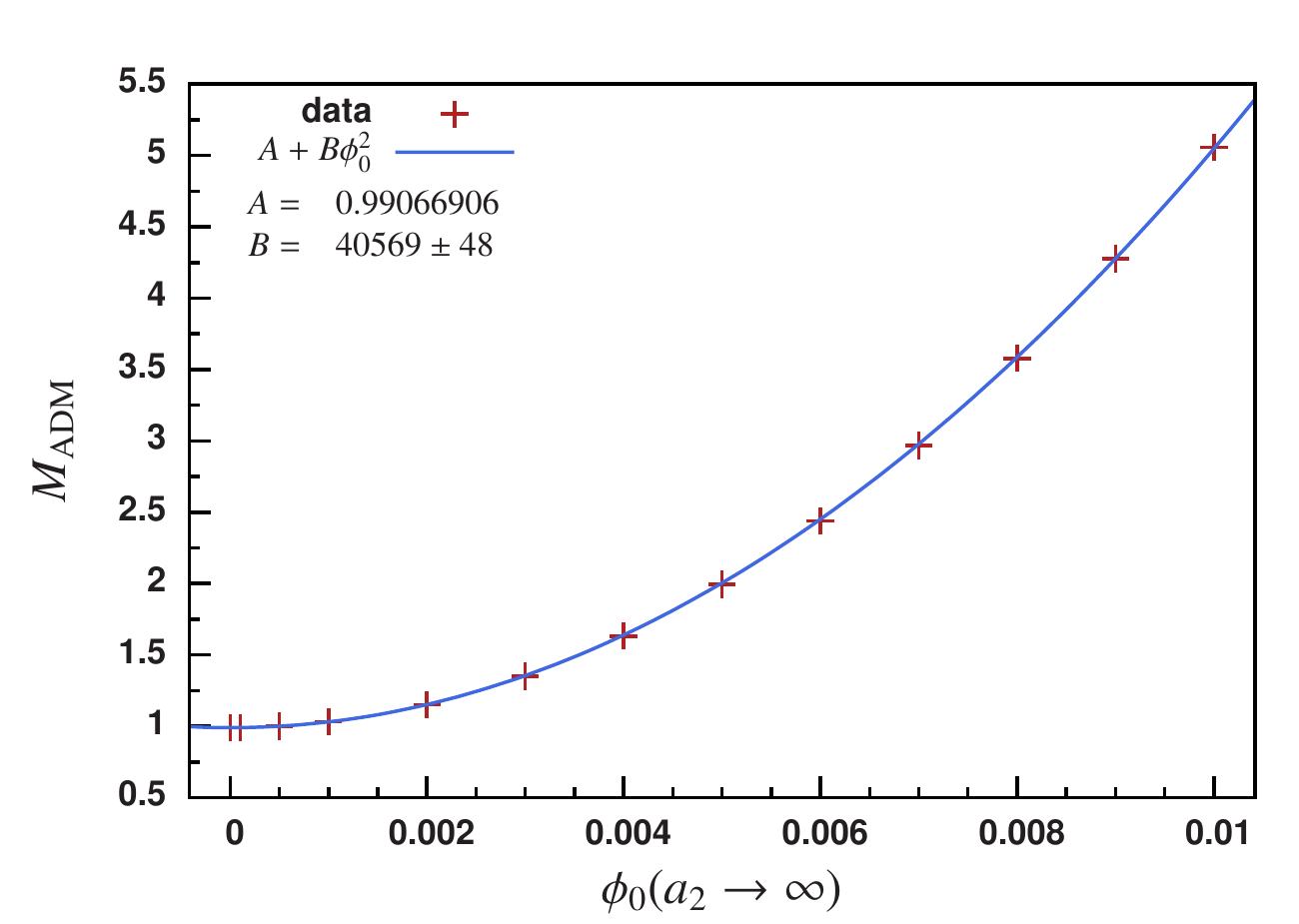}
  \caption{ADM  mass   $\MADM$  as  function  of   $\phi_0$  for  $a_2
    \rightarrow \infty$.  The  functional behavior is well represented
    by  a   quadratic  function.}\label{fig:3}
\end{figure}
%|--------------------------------------------------------------------|

%|--------------------------------------------------------------------|
The estimation of the ADM mass gives us a way to analyze the
parameters $\phi_0$ and $a_2$. On one hand, it is possible to
compute $\MADM$ for the case $a_2 \rightarrow \infty$  for several
values of $\phi_0$ (see Table~\ref{tab:1}).  The  result   is  a
quadratic  relationship  (see figure~\ref{fig:3}).   The quadratic
behavior  is consistent  with the fact that the coefficient of
$\psi_0$ in \eqref{eq:31} for the scalar field profile \eqref{eq:37}
is quadratic in the amplitude $\phi_0$.
%|--------------------------------------------------------------------|
%|--------------------------------------------------------------------|
\begin{figure}[tbp]
  \centering
  \includegraphics[width=85mm]{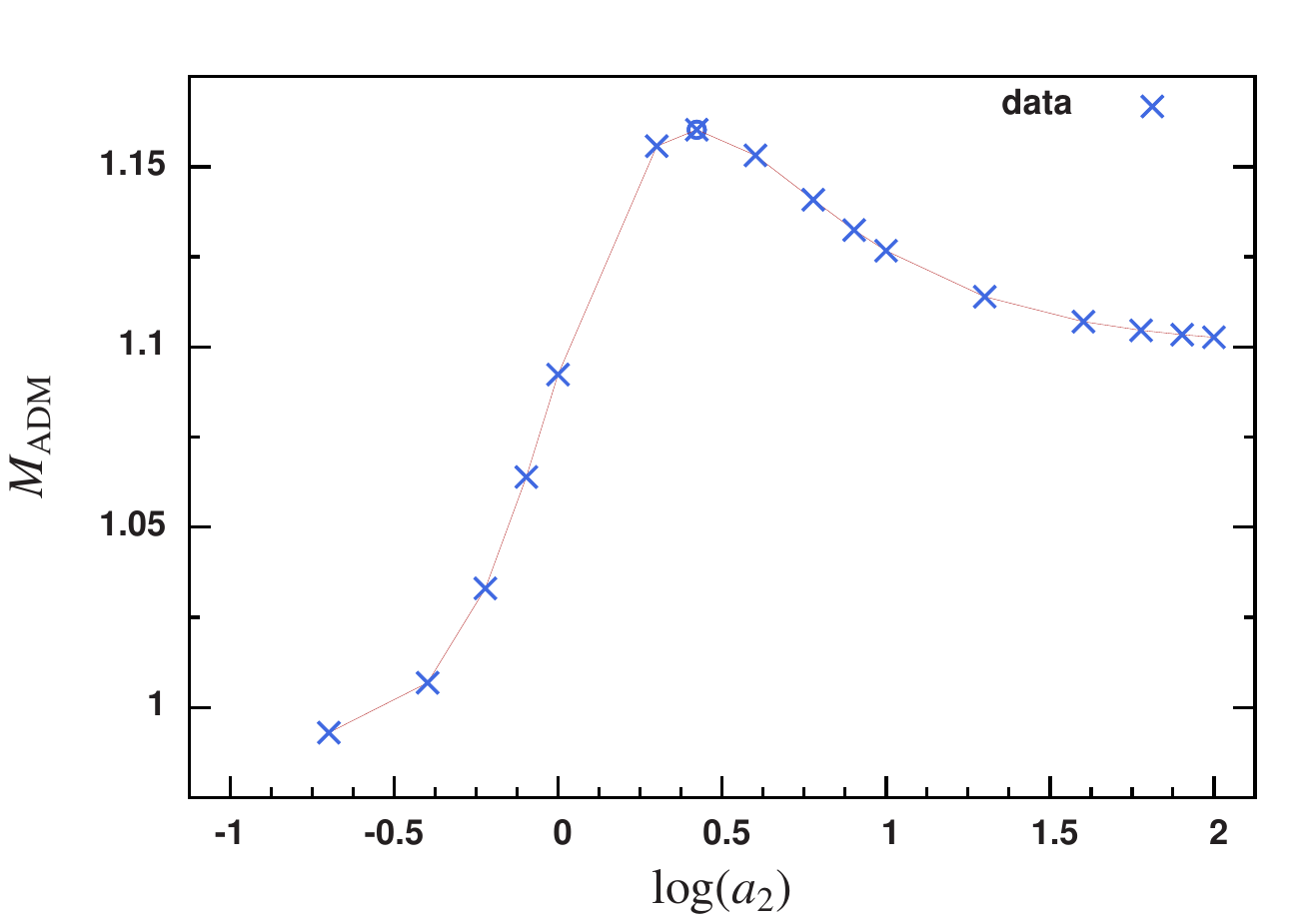}
  \caption{ADM mass $\MADM$ as  function of $\log(a_2)$. The amplitude
    of  the  scalar  field  is $\phi_0=0.001642$.  The  cross-circle
    symbol  denotes the  maximum value  $\MADM=1.16023966$  located at
    $a_2=2.64353$.   The   value  of  $a_2$  is   estimated  from  the
    maximization of \eqref{eq:38}.}\label{fig:4}
\end{figure}
%|--------------------------------------------------------------------|

%|--------------------------------------------------------------------|
\begin{figure}[tbp]
  \centering
  \includegraphics[width=85mm]{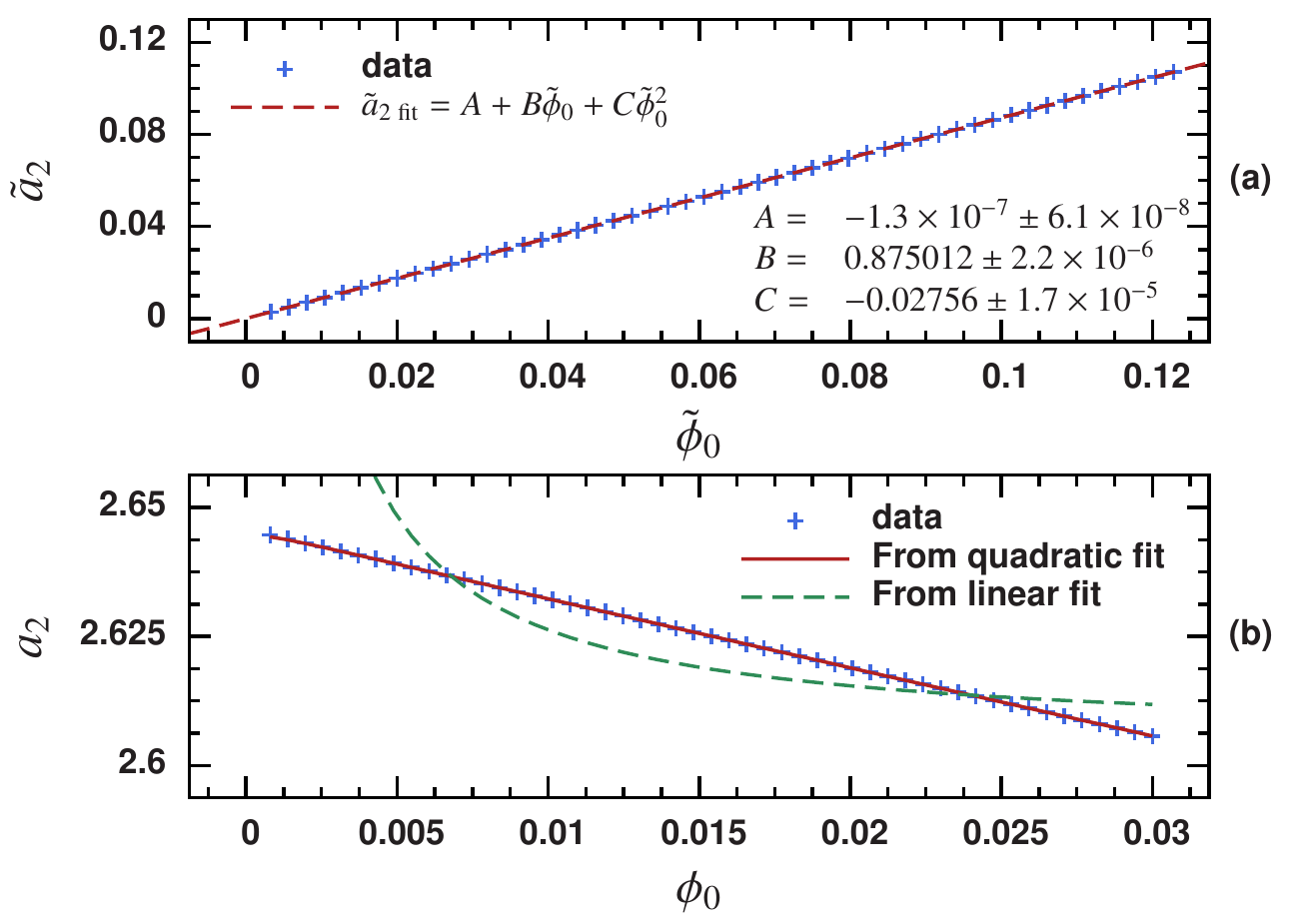}
  \caption{Estimated   values  of   $(\phi_0,  a_2)$   which  maximize
    $\MADM$. The upper  panel \bfa~ shows the result  in the variables
    $(\tilde{\phi}_0,  \tilde{a}_2)$,  where  we  fit a  second  order
    polynomial.  The  lower  panel  \bfb~  shows  the  result  in  the
    variables $(\phi_0, a_2)$. Notice  that in both cases the behavior
    seems to  be linear, however  by using a  linear fit in  the tilde
    variables  the result  does  not  fit the  data  in the  $(\phi_0,
    a_2)$ (dashed line). }\label{fig:5}
\end{figure}
%|--------------------------------------------------------------------|

%|--------------------------------------------------------------------|
On  the other  hand, for  fixed  $\phi_0$, we  analyzed the  functional
behavior of $\MADM$ as function of $a_2$. Figure~\ref{fig:4} shows the
result (in this example $\phi_0=0.001642$).  For this particular value
of $\phi_0$, the ADM mass reaches its maximum value $\MADM=1.16023966$
when $a_2=2.64353$.  The estimation of  the value $a_2$ comes from the
maximization  of  the product  of  the  coefficients  of $\psi_0$  and
$\psi_0^5$ (see right hand side of \eqref{eq:31}):
%|--------------------------------------------------------------------|
\begin{align}
\mathcal{C}=&\sqrt{(\tilde{\phi}_0-\tilde{a}_2)\tilde{a}_2^3}(1-e^{\tilde{a}_2})^2e^{-2\tilde{a}_2}\label{eq:38} \\
\sim& \phi'(r=r_0+\sqrt{\sigma/2})^2 V(r=r_0)\nonumber
\end{align}
%|--------------------------------------------------------------------|
where    we   define   $\tilde{\phi}_0:=4\sqrt{\pi/3}    \phi_0$   and
$\tilde{a}_2:=\tilde{\phi}_0  a_2^2 /  (a_2^2+1)$.   Notice that  with
respect to the radial coordinate $r$ the coefficients are evaluated in
their   respective  maximums.    We   are  looking   for  the   values
$(\phi_0,a_2)$ which maximize the product instead of the maximum value
of  $\mathcal{C}$.   Therefore, we  can  drop  all the  multiplicative
constants.   The  maximization  of  $\mathcal{C}$  is  performed  with
respect to  the variable $\tilde{a}_2$.   The extrema of  the function
reduces to computing the roots of
%|--------------------------------------------------------------------|
\begin{equation}
\mathcal{C}'(\tilde{a}_2) \sim 4 \tilde{a}_2 (\tilde{a}_2+\e^{\tilde{a}_2}+\tilde{\phi}_0 -1)-3(\e^{\tilde{a}_2}-1)\tilde{\phi_0}.
\end{equation}
%|--------------------------------------------------------------------|
We  computed   the  values  numerically   using  \textsc{Mathematica}.
Figure~\ref{fig:5}  shows the  result.   From the  numerical data,  it
appears that  $\tilde{a}_2$ is  a linear function  of $\tilde{\phi}_0$
(see Figure~\ref{fig:5}-\bfa). However, a  comparison of the data with
the fitted linear function showed us that a higher order polynomial is
better  a approximations. We  choose a  second order  polynomial since
higher order polynomials do not exhibit a significant reduction of the
errors.   The results  for $(\phi_0,  a_2)$ variables  confirm  that a
quadratic     fit      is     a     better      approximation     (see
Figure~\ref{fig:5}-\bfb). Note that  in the interval investigated $a_2
\sim  2.64$.   In  international   system  units,  it  corresponds  to
$10^{11}$m${}^2$ (considering  the typical gravitational  wave sources
of  BBH  for LIGO).   This  value maximizes  the  \fR  effect for  BBH
collisions.
%|--------------------------------------------------------------------|

\section{Evolution of equal mass binary black holes in \fR theory}
\label{sec:evolution-equal-mass}

\subsection{Numerical method}
\label{sec:numerical-method-2}

%|--------------------------------------------------------------------|
The evolution of  the black hole and scalar field  is solved using the
\AMSSNCKU                           code                          (see
\cite{CaoYoYu08,GalBruCao10a,CaoLiu11,Cao12,CaoHil12}).        Although
\AMSSNCKU code supports both vertex center and cell center grid style,
we use the cell center  style.  We use finite difference approximation
of 4th  order.  We  update the  code to include  the dynamics  of real
scalar field  equations \eqref{eq:17}  and \eqref{eq:18}.  We  use the
outgoing radiation boundary condition for all variables.  In addition,
we update  our code  to support  a combination of  box and  shell grid
structures (according to \cite{Tho04,PolReiSch11}).
%|--------------------------------------------------------------------|

%|--------------------------------------------------------------------|
The numerical  grid consists of  a hierarchy of nested  Cartesian grid
boxes and a shell which includes six coordinate patches with spherical
coordinates   ($\rho,\sigma,r$).    For   symmetric  spacetimes,   the
corresponding symmetric  patches are dropped.   Particularly, we adopt
equatorial symmetry.   For the nested  Cartesian grid boxes,  a moving
box  mesh refinement is  used.  For  the outer  shell part,  the local
coordinates  of the  six shell  patches are  related to  the Cartesian
coordinates by
%|--------------------------------------------------------------------|
\begin{align}
  \pm x & &\text{patch:} \quad \rho = \arctan(y/x), \sigma = \arctan(z/x), \\
  \pm y & &\text{patch:} \quad \rho = \arctan(x/y), \sigma = \arctan(z/y), \\
  \pm z & &\text{patch:} \quad \rho = \arctan(x/z), \sigma = \arctan(y/z),
\end{align}
% |--------------------------------------------------------------------|
where both angles  ($\rho,\sigma$) range over $(-\pi/4:\pi/4)$.
% |--------------------------------------------------------------------|

% |--------------------------------------------------------------------|
Notice  that  positive  and  negative Cartesian  patches  are
related through the same coordinate transformation.  This coordinate
choice is right  handed in $+x$,  $-y$, $+z$  patches and  left
handed  in $-x$, $+y$,   $-z$  patches.    Disregarding   parity
issues,   left-handed coordinates  do not  bring us  any
inconvenience.  We have applied this coordinate choice to
characteristic evolutions in \cite{Cao13}. For an alternative
approach, see \cite{Tho04, PolReiSch11}.
%|--------------------------------------------------------------------|
The coordinate  radius $r$ relates to the  global Cartesian coordinate
through
%|--------------------------------------------------------------------|
\begin{align}
r=\sqrt{x^2+y^2+z^2}.
\end{align}
%|--------------------------------------------------------------------|
All  dynamical equations for  numerical evolution  are written  in the
global Cartesian coordinate. The local coordinates ($\rho,\sigma,r$) of
the six  shell patches  are used to  define the numerical  grid points
with  which the  finite  difference is  implemented.  The  derivatives
involved   in  the   dynamical   equations  in   the  Cartesian   grid
$x^i=(x,y,z)$ are  related to the  spherical derivatives in  the shell
coordinates $r^i=(\rho,\sigma,r)$ through
%|--------------------------------------------------------------------|
\begin{align}
  \frac{\partial}{\partial x^i}
  &=
  \left(\frac{\partial r^j}{\partial x^i}\right) \frac{\partial}{\partial r^i}, \label{eq:39}\\
  \frac{\partial^2}{\partial x^i\partial x^j}
  &=
  \left(\frac{\partial r^k}{\partial x^i} \frac{\partial r^l}{\partial x^j}\right) \frac{\partial^2}{\partial r^k \partial r^l} +
  \left(\frac{\partial^2 r^k}{\partial x^i\partial x^j}\right) \frac{\partial}{\partial r^k}.\label{eq:40}
\end{align}
% |--------------------------------------------------------------------|
The  spherical  derivatives  in  \eqref{eq:39} and  \eqref{eq:40}  are
approximated by center finite difference.
%|--------------------------------------------------------------------|

%|--------------------------------------------------------------------|
In the  spherical shell two  patches share a common  radial coordinate
and adjacent patches share the angular coordinate perpendicular to the
mutual boundary. Therefore,  it is not necessary to  perform a full 3D
interpolation between the overlapping shell ghost zones.  Moreover, it
is enough to perform a  1D interpolation parallel to the boundary (see
\cite{Tho04,HilBerThi12} for  details).  For this purpose,  we use 5th
order  Lagrangian  interpolation   with  the  most  centered  possible
stencil.
%|--------------------------------------------------------------------|

%|--------------------------------------------------------------------|
For the  interpolation between shells and the  coarsest Cartesian grid
box,  we  use  a 5th  order  Lagrange  interpolation.   This is  a  3D
interpolation  done through three  directions successively.   The grid
structure for  boxes and shells  are different.  There is  no parallel
coordinate  line between the  grid structures.   Therefore, we  have a
region  which  is double  covered.   Similar  to  the mesh  refinement
interface, we  also use six buffer  points in the box  and shell.  The
buffer points are  re-populated at a full Runge-Kutta  time step.  For
parallelization, we  split the shell patches  into several sub-domains
in three directions.  The same is done for boxes.
%|--------------------------------------------------------------------|

\begin{figure}[tbp]
  \centering
  \includegraphics[width=85mm]{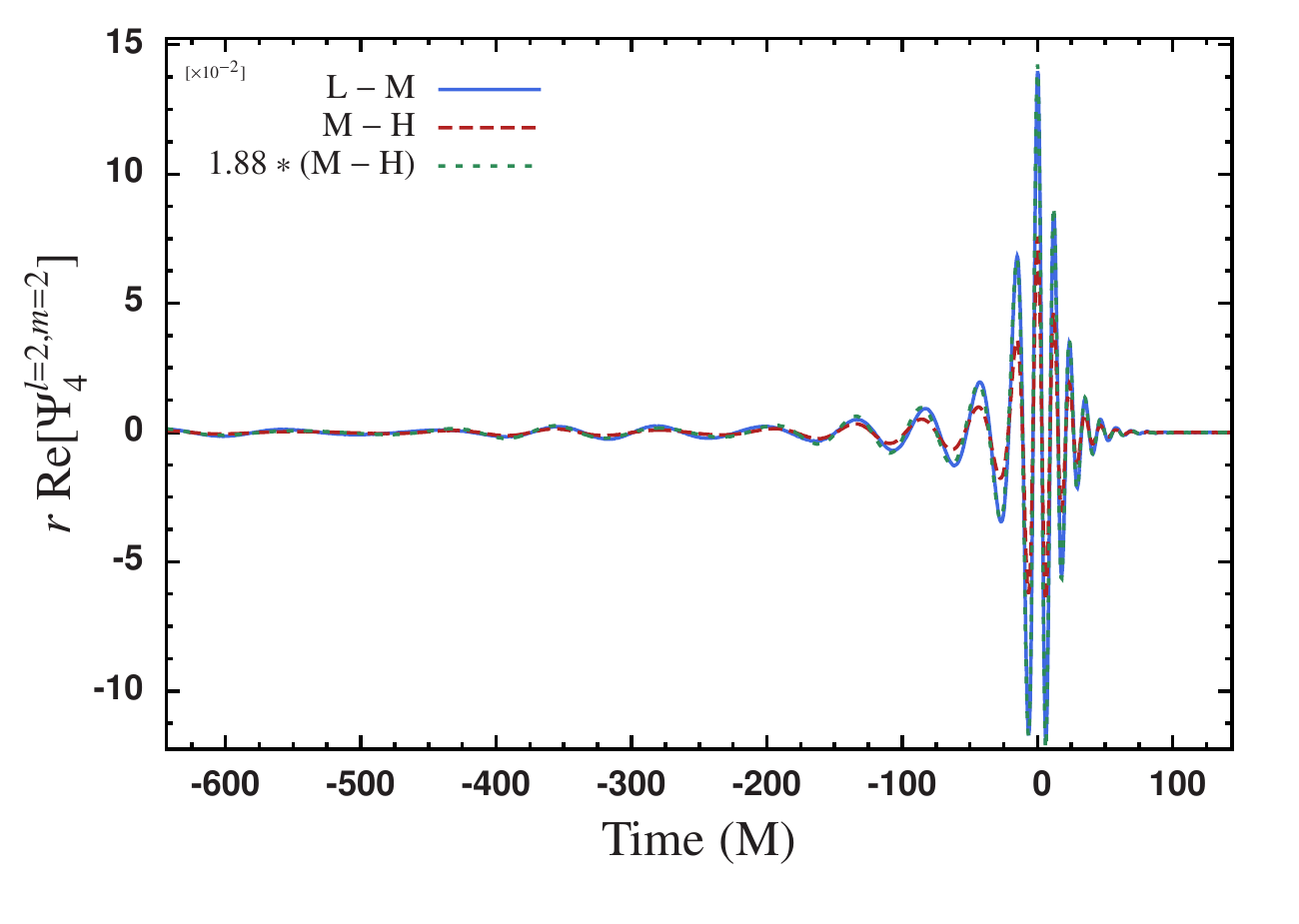}
  \caption{Convergence test  of the  waveform. Real part  of $\ell=2$,
    $m=2$  mode  of  $\Psi_4$.    The  evolution  corresponds  to  the
    parameters     $a_2=2.64418$     and    $\phi_0=0.000959$     (see
    Table~\ref{tab:3}). The plot shows the differences between the low
    (L) and  medium (M) resolutions  (solid line), and the  medium (M)
    and high  (H) resolutions  (dashed line).  The  difference between
    the medium  and high  is scaled by  1.88 which corresponds  to 4th
    order convergence (dotted line).   The corresponding values of the
    grid  size for  the finest  refinement level  are (L)  0.009 $M$, (M)
    0.0079 $M$ and (H) 0.007 $M$.}\label{fig:6}
\end{figure}

%|--------------------------------------------------------------------|
We have tested the convergence behavior of the updated \AMSSNCKU code.
Fig.~\ref{fig:6}, shows the  waveform produced with three resolutions.
The corresponding  values of the  grid size for the  finest refinement
level are 0.009 $M$, 0.0079 $M$ and 0.007 $M$.  From here-on, we refer
to these  values as the low  (L), medium (M) and  high (H) resolutions
respectively.  We  shift the time in  order to align  the waveforms at
the  maximum amplitude  of  $\Psi_{4,22}$.  The  results presented  in
sections  \ref{sec:results-2}  and  \ref{sec:results-3} are  performed
with the medium resolution.
%|--------------------------------------------------------------------|

%|--------------------------------------------------------------------|
%|--------------------------------------------------------------------|

\begin{figure}[tbp]
  \centering
  \includegraphics[width=85mm]{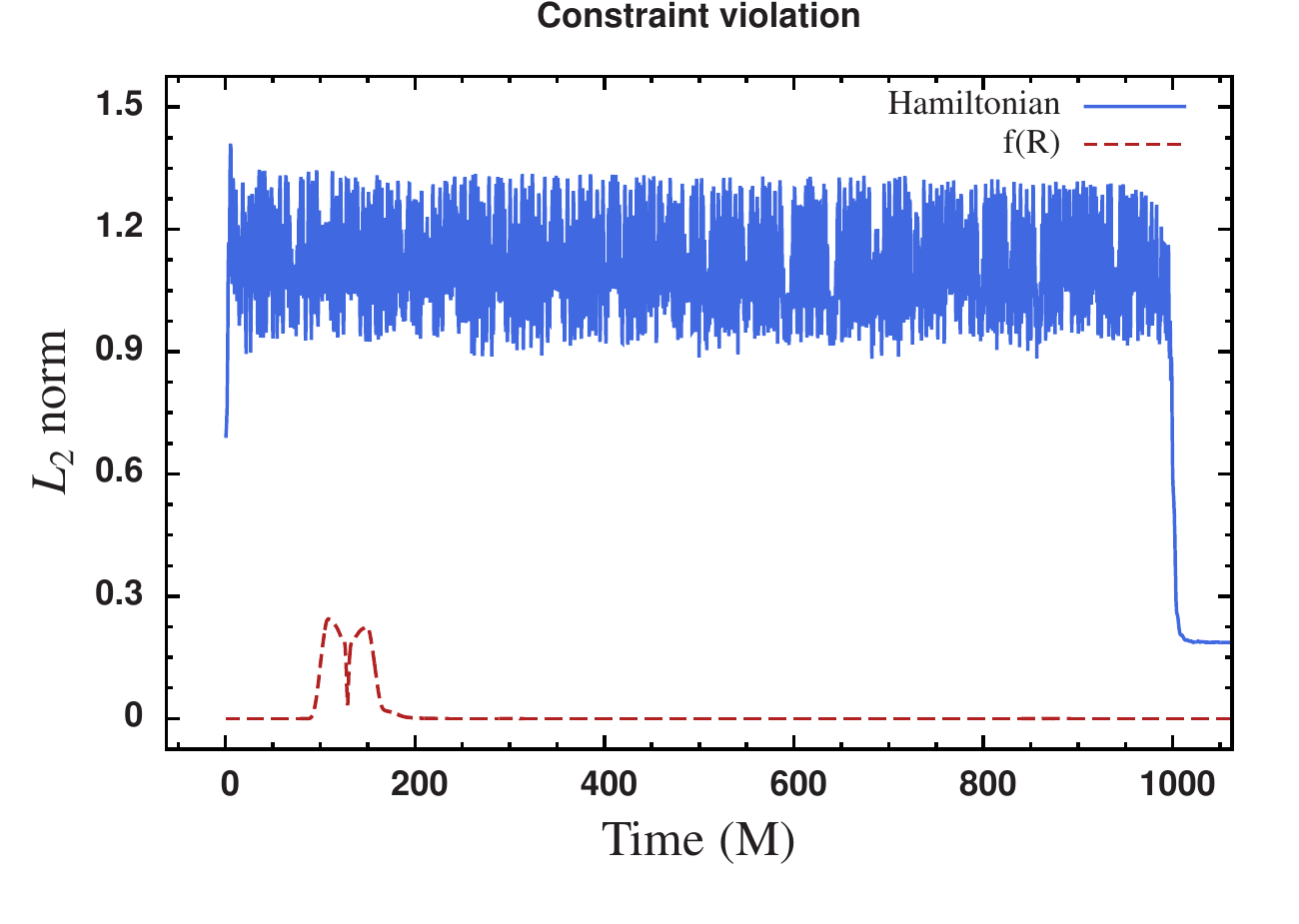}
  \caption{L2  norm  of   Hamiltonian  constraint  violation  and  \fR
    constraint  violation  \eqref{eq:41}.  Here, $a_2  \rightarrow
    \infty$ and $\phi_0=0.000959$.}\label{fig:7}
\end{figure}

%|--------------------------------------------------------------------|
The equation \eqref{eq:14} represents a constraint equation which is introduced
by reducing the 4th order  derivative dynamical formulation to the 2nd
order. Based on 3+1 formalism, we have
%|--------------------------------------------------------------------|
\begin{align}
{}^{(4)}R=-2\DLie_nK+R+K^2+K_{ij}K^{ij}-\frac{2}{\alpha}D_iD^i\alpha.
\end{align}
%|--------------------------------------------------------------------|
Substituting  $\DLie_nK$ with  the evolution  equations  for
$K_{ij}$ results in
%|--------------------------------------------------------------------|
\begin{align}
{}^{(4)}R&=8\pi(3E-S)-R-K^2+K_{ij}K^{ij}\\
&=16\pi(D_i\phi D^i\phi+3V)-R-K^2+K_{ij}K^{ij}.
\end{align}
%|--------------------------------------------------------------------|
Therefore, the constraint equation reads as
%|--------------------------------------------------------------------|
\begin{align}
&\ln{\left(1+2a_2[16\pi(D_i\phi
D^i\phi+3V)-R-K^2+K_{ij}K^{ij}]\right)}\nonumber\\
&=\frac{4\sqrt{\pi}}{\sqrt{3}}\phi.\label{eq:41}
\end{align}
%|--------------------------------------------------------------------|
From here-on,  we will refer  to \eqref{eq:41} as the  \fR constraint.
In  Fig.~\ref{fig:7}, we  show an  example  of the  violation of  this
constraint during our simulations. This violation of \fR constraint is
much smaller than that of the Hamiltonian constraint.
%|--------------------------------------------------------------------|

%|--------------------------------------------------------------------|
%|--------------------------------------------------------------------|

\subsection{Initial scalar field setup}
\label{sec:results-2}

\begin{figure}[tbp]
  \centering
  \includegraphics[width=85mm]{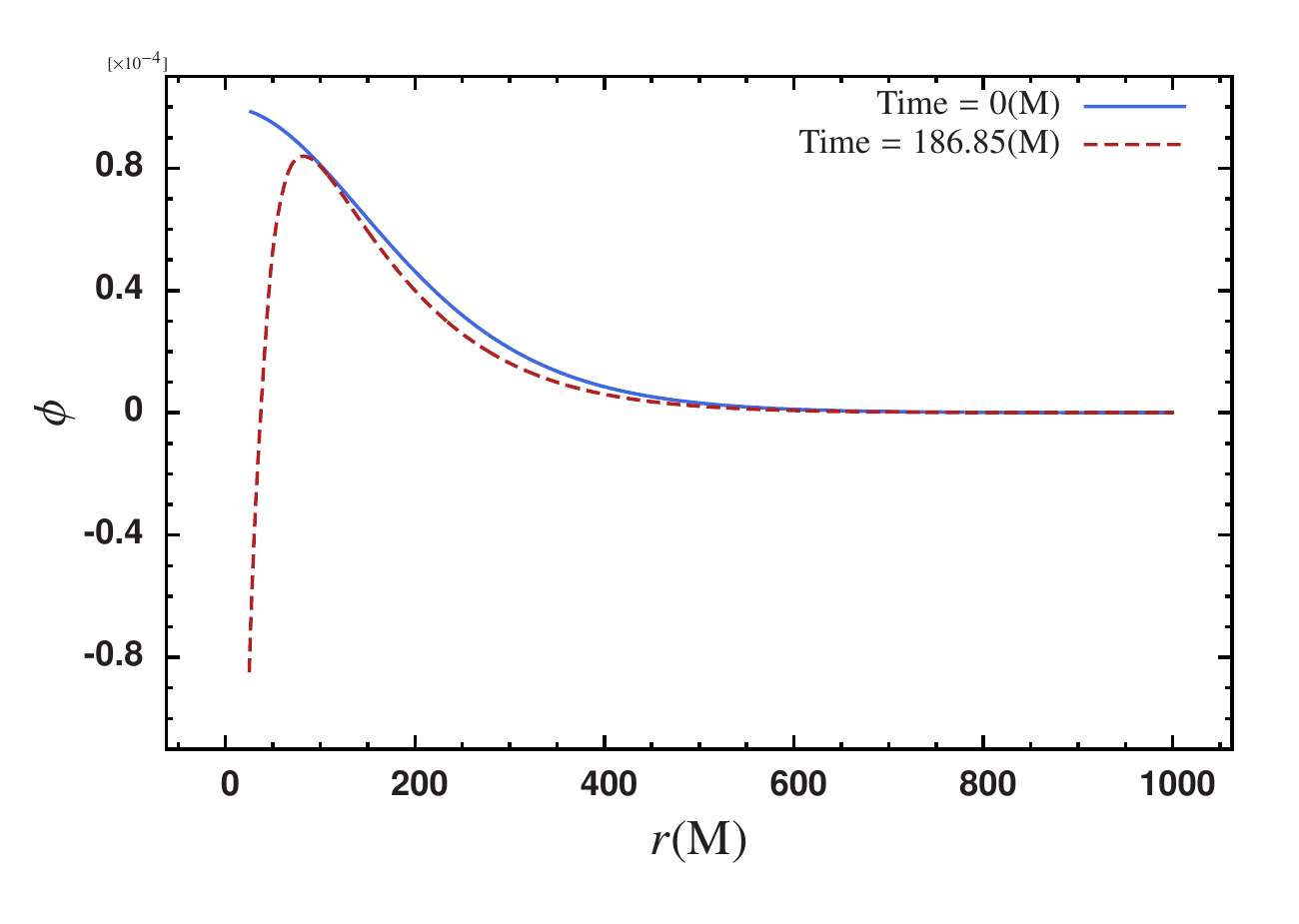}
  \caption{Snapshots  of   a  scalar   field  evolving  with   a  BBH.
    $\mathrm{Time}=0$  M  corresponds  to  the  ground  state  of  the
    Schr\"odinger-Newton  system  considered  in~\cite{GuzUre04}.   At
    $\mathrm{time}=186.85$ M, a shell shape forms.}\label{fig:8}
\end{figure}

%|--------------------------------------------------------------------|
One way  to interpret \fR theory  is as an effective  model of quantum
gravity. In  the astrophysical context,  it is natural to  assume that
the  systems are  in  their ground  states,  and correspondingly,  the
scalar  field takes the  profile of  the ground  state of  the related
quantum gravity system.
%|--------------------------------------------------------------------|
We simulate the development  of the scalar field from the ground
state     of     the     Schr\"odinger-Newton    system     considered
in~\cite{GuzUre04}.    Other   authors   model   the   dark   matter
halo~\cite{MagMat12} in the center of  a galaxy with a similar profile
(see e.g.,~\cite{SalBor03}).   Our result shows that  the scalar field
evolves from  the ground state  configuration to a  shell-type profile
(similar to  \eqref{eq:37}).  Moreover, the  shell forms in  the early
stages  of the  evolution. Fig.~\ref{fig:8}  shows two  snapshots, the
initial ground state profile and the final shell configuration.
%|--------------------------------------------------------------------|
In our test,  the initial profile of the scalar  field is some general
Gaussian shape, and the shell shape soon forms. Our results imply that
the formation of a shell shape is generic in coupled systems of scalar
field and BBH.
%|--------------------------------------------------------------------|

%|--------------------------------------------------------------------|
%|--------------------------------------------------------------------|

%|--------------------------------------------------------------------|
Considering  the development  of a  scalar  field shell  in the  early
stages of the  formation of a BBH system,  we starte the evolution
with the profile \eqref{eq:37}. The parameters used in our simulations
are listed in Table~\ref{tab:3}.   We divide the parameters into three
groups.  The  first group, $a_2=0$, $\phi_0=0$  corresponds to general
relativity.  The second group, $a_2 \rightarrow \infty$ corresponds to
the free  EKG equations.  In  this case, the  scalar field in  the far
zone is weak. Therefore, the waveforms in the Jordan frame are similar
to   the  waveforms  in   the  Einstein   frame.   The   third  group,
$0<a_2<\infty$ corresponds  to general  $f(R)$ theory.  In  this case,
the value $a_2$ is the one which maximizes $\MADM$ for given $\phi_0$.
%|--------------------------------------------------------------------|
\begin{table}[btp]
  \begin{ruledtabular}
    \caption{Parameters of  the scalar field.  There  are three groups
      of parameters.  $a_2=0$  corresponds to general relativity; $a_2
      \rightarrow \infty$  group corresponds  to the EKG  equations in
      general  relativity; and  $0<a_2<\infty$ corresponds  to general
      \fR theory.}\label{tab:3}
    \begin{tabular}{l|cc}
       $\MADM$  &  $\phi_0$  &  $a_2$ \\ \hline
      % |-------------------------------------------------------------|
       0.99067&     0    &      0\\ \hline
       0.99062&     0.000048&   $\infty$\\
       0.99980&     0.000480&   $\infty$\\
       1.02756&     0.000959&   $\infty$\\ \hline
       0.99067&     0.000048&   2.61877\\
       1.00490&     0.000480&   2.64297\\
       1.04790&     0.000959&   2.64418
      %|--------------------------------------------------------------|
    \end{tabular}
  \end{ruledtabular}
\end{table}
%|--------------------------------------------------------------------|

\subsection{Results}\label{sec:results-3}

%|--------------------------------------------------------------------|
In this  subsection, we present  the numerical simulation  results for
the BBH evolution  in \fR theory.  We focus  on the comparison between
\fR and GR evolution.  We refer  to the difference between them as the
\fR effect.
%|--------------------------------------------------------------------|

\subsubsection{Dynamics of the scalar field induced by binary black holes}
\label{sec:dynamic-sf-bbh}

%|--------------------------------------------------------------------|
\begin{figure}[tbp]
  \centering
  \includegraphics[width=85mm]{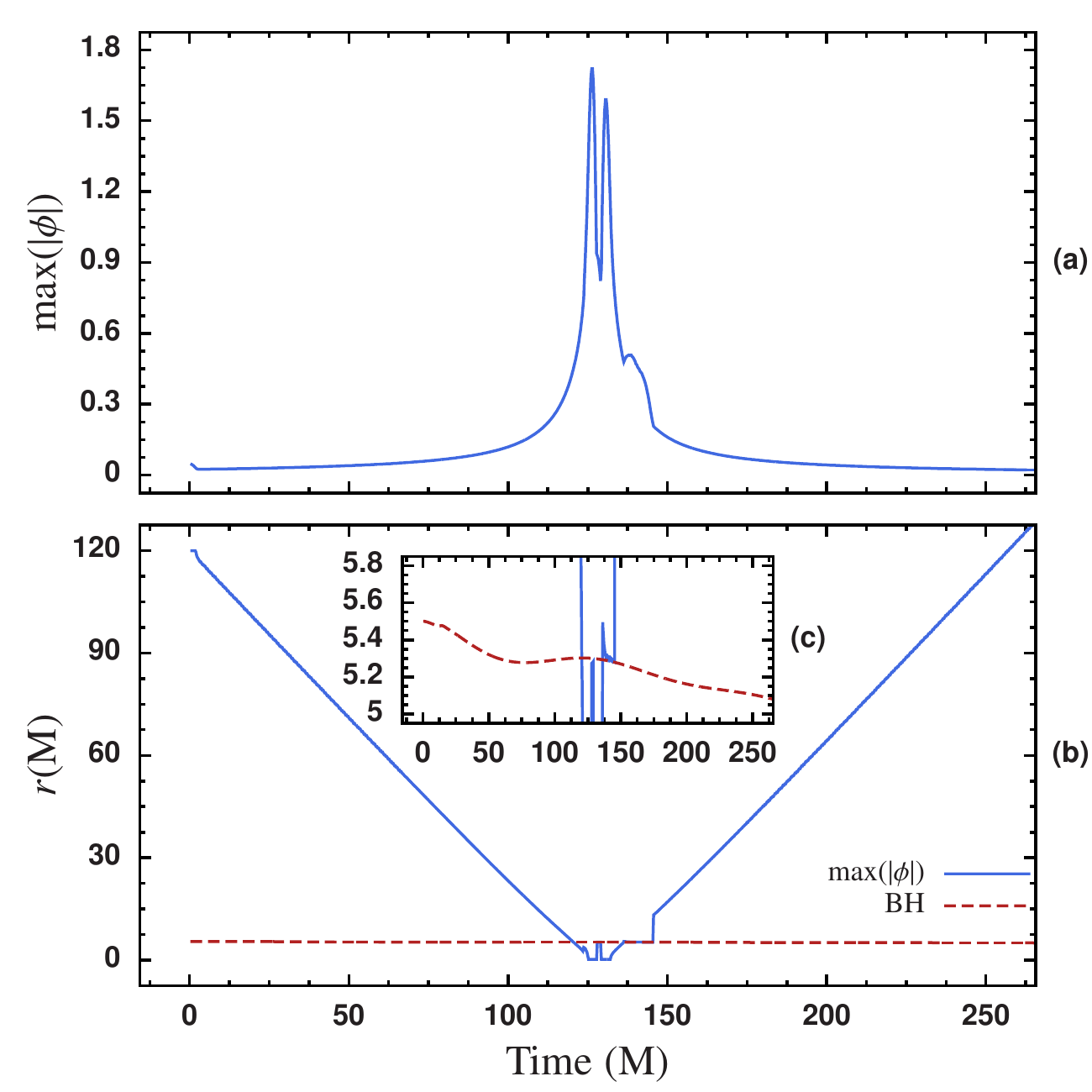}
  \caption{Dynamics of scalar field induced by BBH. The parameters are
    $a_2    \rightarrow    \infty$    and    $\phi_0=0.000959$    (see
    Table~\ref{tab:3}).  The  upper panel  \bfa~ shows the  maximum of
    $|\phi|$ as  a function of  time.  The external lower  panel \bfb~
    shows the radius position of  one black hole and the corresponding
    radius  position of  the maximum  of the  scalar  field.  Internal
    lower panel \bfc~ shows the magnification of the collision part of
    the scalar field and the black hole.}\label{fig:9}
\end{figure}
%|--------------------------------------------------------------------|

%|--------------------------------------------------------------------|
The characteristic dynamics of the  scalar field in our simulations is
the  following.   Starting  from  a  shell  shape,  the  scalar  field
collapses towards  the central BBH.   Then, the maximum of  the scalar
field reaches  the black  holes.  At that  moment in the  evolution, a
burst of gravitational radiation is  produced . After that, the scalar
field continues collapsing towards the origin of the numerical domain.
The  BBH  excites the  surrounding  scalar  field.  The  perturbations
produced by the BBH collapses  to the origin, thereby joining the main
part of the scalar field.  After reaching the origin, the scalar field
is scattered  in the outgoing  direction. Once the scalar  field moves
outside of the orbit of the BBH,  it is attracted by the BBH again and
remains there for  some time. The scalar field  slowly radiates to the
exterior of the  numerical domain. In the process,  part of the scalar
field is absorbed by the black holes.
%|--------------------------------------------------------------------|

%|--------------------------------------------------------------------|
In Fig.~\ref{fig:9}-\bfa~ we show the maximum of $|\phi|$ with respect
to time.  Since  the scalar field approximates a  shell shape, we only
consider  the radial  position.  The  change in  the amplitude  of the
scalar  field represents  the  collapsing stage  (increments) and  the
scattering  stage  (decrements).   There  are two  main  peaks  around
$\mathrm{time}=125$  M.  The  first  peak corresponds  to the  initial
collapse (before reaching the  BBH).  The second peak corresponds with
the excitation of the scalar field  produced by the BBH. A small third
peak corresponds to the attraction produced by the BBH.
%|--------------------------------------------------------------------|

%|--------------------------------------------------------------------|
Fig.~\ref{fig:9}-\bfb~ shows the  radal position of max($|\phi|$) with
respect to  time (solid  line) and the  radial position of  black hole
(dashed  line).   The  main   collapsing  and  scattering  process  is
clear. There  are four coincidences of  the scalar field  and the BBH.
Three of them correspond to the peaks showed in Fig.~\ref{fig:9}-\bfa.
We enlarge the detail of the encounters in Fig.~\ref{fig:9}-\bfc.
%|--------------------------------------------------------------------|

\begin{figure}[tbp]
  \centering
  \includegraphics[width=85mm]{\imgdir/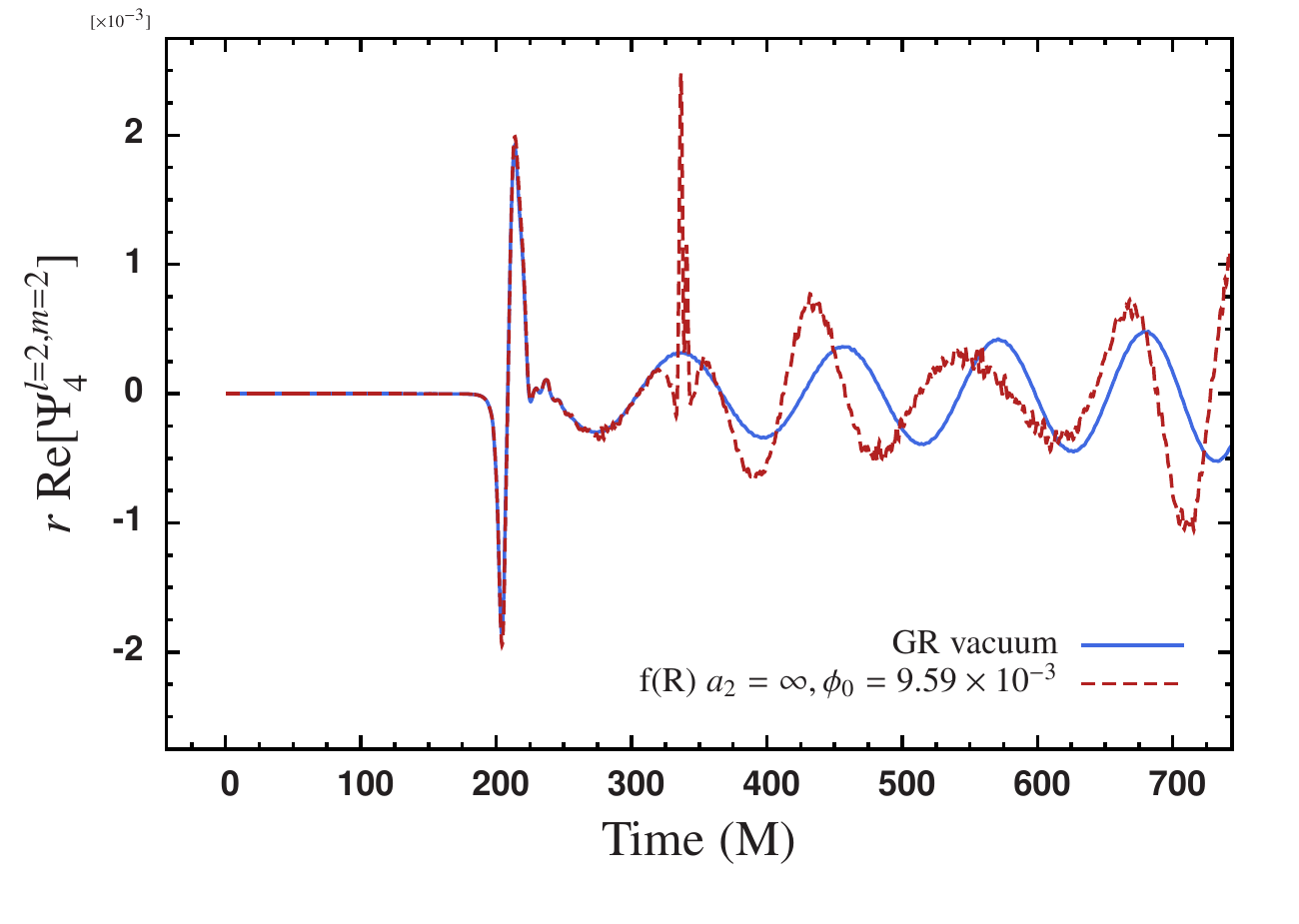}
  \caption{Comparison of  the initial part  of the waveform for  a BBH
    collision in  GR and \fR  theory with parameters  $a_2 \rightarrow
    \infty$  and $\phi_0=0.000959$. The  collision between  the scalar
    field and the  BBH produces a burst of  gravitational radiation at
    roughly $\mathrm{time}=340$ M.  }\label{fig:10}
\end{figure}

%|--------------------------------------------------------------------|
As mentioned above, the collision between the scalar field and the BBH
produces a  burst of gravitational  radiation. Fig.~\ref{fig:10} shows
the corresponding waveform of the evolution presented previously (with
parameters $a_2  \rightarrow \infty$ and  $\phi_0=0.000959$).  In this
plot, we extract the waves at $r=200$ M.  After the radiation produced
by the initial data configuration (so-call junk radiation), there is a
peak  at about  $\mathrm{time}=340$ M  (dashed line).   This  burst of
radiation is not present in  the BBH case (solid line).  Moreover, the
pattern is encoded in every even $m$ mode of $\Psi_4$.
%|--------------------------------------------------------------------|

\begin{figure}[tbp]
  \centering
  \includegraphics[width=85mm]{\imgdir/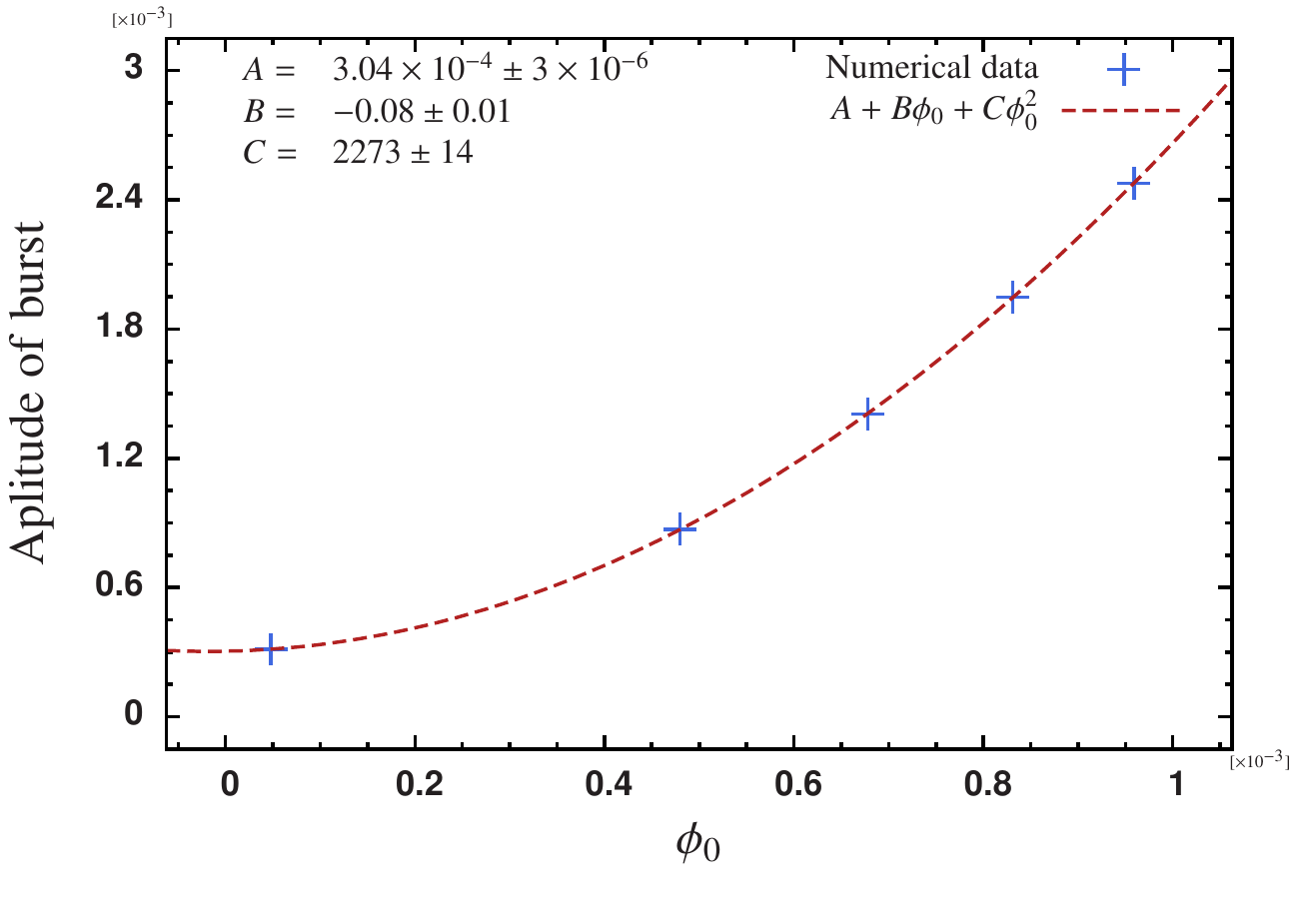}
  \caption{Burst  amplitude  as  a  function  of  the  initial  scalar
    parameter $\phi_0$.   In this case $a_2  \rightarrow \infty$.  The
    fitting  parameters   are  $A=  3.04\times   10^{-4}  \pm  3\times
    10^{-6}$, $B=-0.08 \pm 0.01$ and  $C=2273 \pm 14$. Notice that the
    value  of $A$  is  approximately  equal to  the  amplitude of  the
    waveform for GR. }\label{fig:11}
\end{figure}

%|--------------------------------------------------------------------|
Fig.~\ref{fig:11} shows  the dependence of the amplitude  of the burst
as a function  of $\phi_0$. The functional behavior  is well represented
by a  quadratic function $A+B \phi_0+C \phi_0^2$,  with $A= 3.04\times
10^{-4} \pm 3\times 10^{-6}$, $B=-0.08 \pm 0.01$ and $C=2273 \pm 14$.
%|--------------------------------------------------------------------|
%|--------------------------------------------------------------------|

%|--------------------------------------------------------------------|
%|--------------------------------------------------------------------|

\begin{figure}[tbp]
  \centering
  \includegraphics[width=85mm]{\imgdir/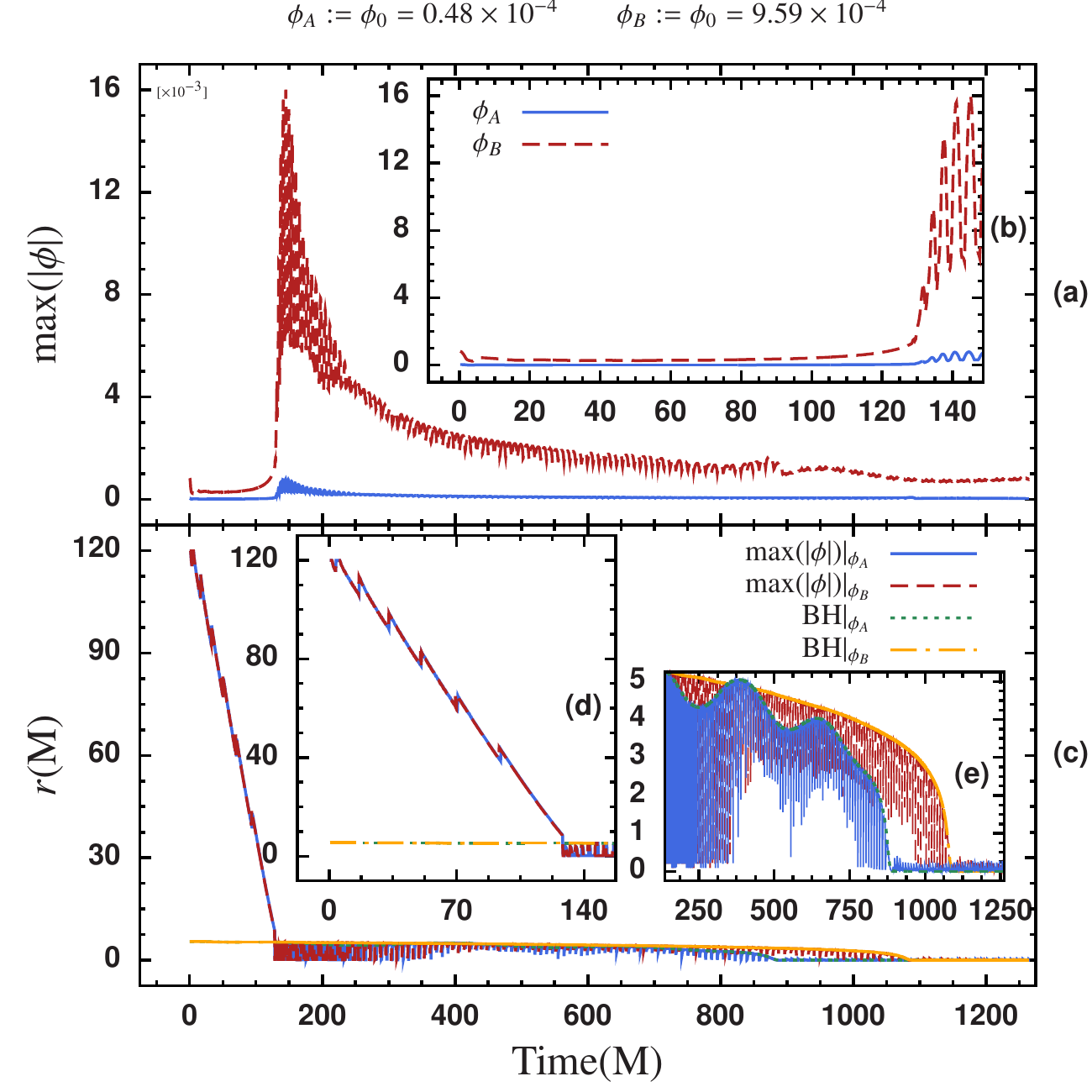}
 \caption{ Dynamics of scalar field induced by BBH. The parameters are
   \{$\phi_0=0.000048$,     $a_2=2.61877$\}    (solid     line)    and
   \{$\phi_0=0.000959$,  $a_2=2.64418$\}  (dashed  line).   The  upper
   panel  ~\bfa~  shows the  maximum  of  $\vert  \phi_0 \vert$  as  a
   function  of  time.   The  internal  upper  panel  ~\bfb~  shows  a
   magnification of  the initial  evolution. The external  lower panel
   ~\bfc~ shows the  radius positions of one black  hole for each case
   (dotted  and  dash-dotted   lines)  and  the  corresponding  radius
   positions of the maximum of  the scalar field.  Internal lower panel
   ~\bfd~ shows a  magnification of the collapse of  the scalar field.
   Internal  lower panel  \textbf{(e)}  shows a  magnification of  the
   merger  phase.   Notice that  in  this  case  the scalar  field  is
   constantly excited.}\label{fig:12}
\end{figure}

%|--------------------------------------------------------------------|
In the above  description, we have presented the  results for the
free EKG  system ($a_2  \rightarrow \infty$).   For our
representative \fR case, where $a_2$ is finite  but non-vanishing,
the behavior of scalar field   is   qualitatively   different.    We
compared   the   cases $\phi_0=0.000959$   and  $a_2=2.64418$   with
$\phi_0=0.000048$  and $a_2=2.61877$.  Fig.~\ref{fig:12} shows  the
results.  Contrary to the free  EKG  case,  we  found  only one
collapsing  stage  without  the scattering to infinity phase.  In
both cases, almost all of the scalar field was absorbed by the black
holes.  During the collapsing process, the scalar field excites the
spacetime.  The back reaction excites the scalar  field,  thereby
producing several  zig-zags  in its  maximum amplitude  (see Figures
\ref{fig:12}-d).   After the  maximum of  the scalar field passes
over the  black hole, the dynamics of scalar field become much
richer.   The scalar field is constantly  excited near the black
hole.   Fig.\ref{fig:12}-(e) shows  that  the  scalar field  is
trapped in the inner region of  the BBH's orbit.  The black holes
play the role of a semi-reflective boundary. A minor amount of
scalar field escapes to infinity.  In comparison with the free EKG
system, the case $\phi_0=0.000959$  and  finite  $a_2$  introduces a
large  amount  of eccentricity  to  the BBH  system.   However,
there  is no  burst  of gravitational  radiation (which  corresponds
to  the one  presented in Fig.~\ref{fig:10}).
%|--------------------------------------------------------------------|

%|--------------------------------------------------------------------|
%|--------------------------------------------------------------------|
\subsubsection{Dynamics of the binary black hole induced by the scalar
  field}
\label{sec:dynamic-bbh-sf}

%|--------------------------------------------------------------------|
\begin{figure}[tbp]
  \centering \includegraphics[width=85mm]{\imgdir/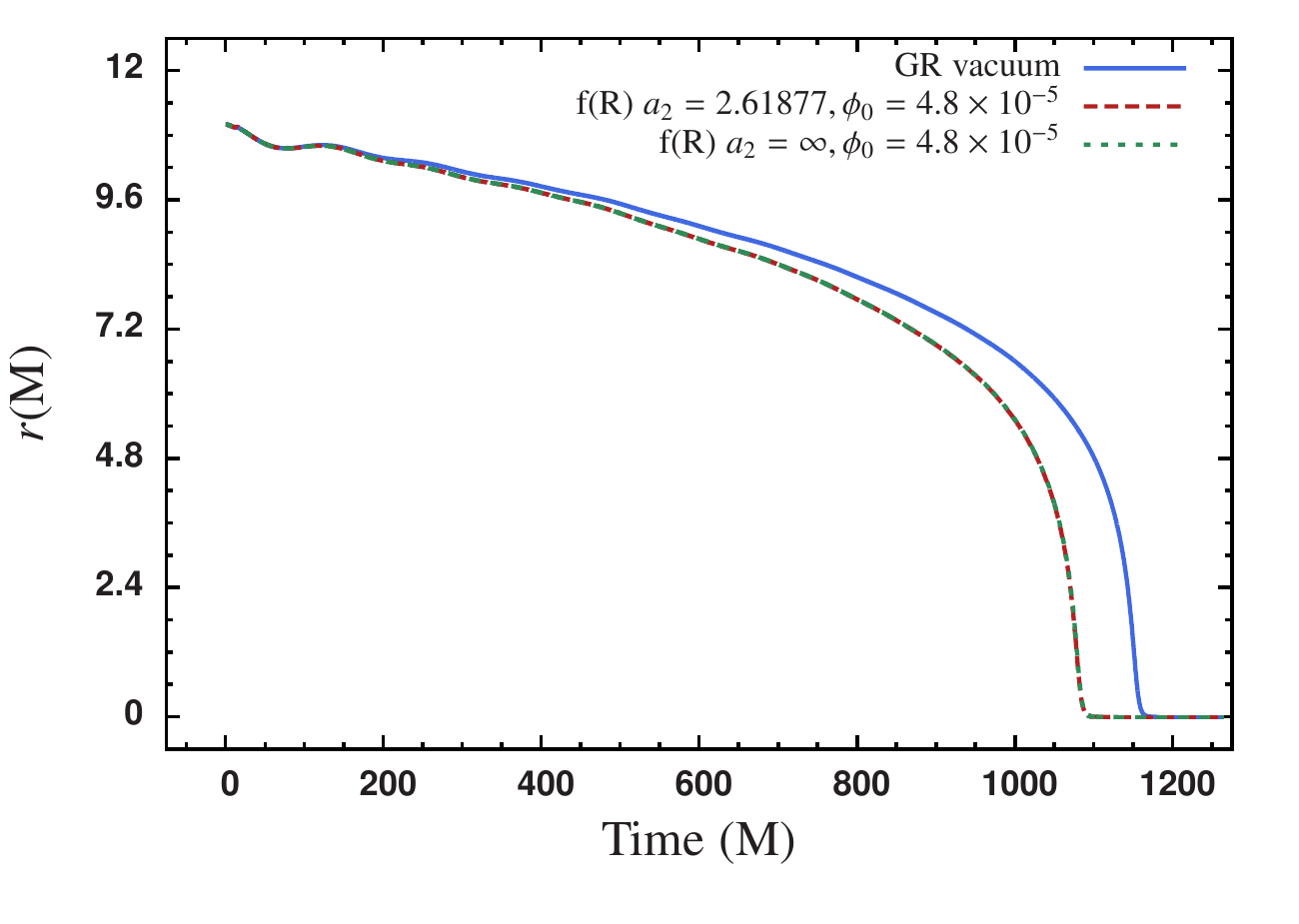}
  \caption{Coordinate separation between  the black holes.  Comparison
    between the GR vacuum case (solid line), a characteristic \fR case
    (dashed line) and the free  EKG case (dotted line). The \fR effect
    makes the  BBH merge  faster than GR  vacuum independently  of the
    total ADM mass.}\label{fig:13}
\end{figure}
%|--------------------------------------------------------------------|

%|--------------------------------------------------------------------|
The trajectory  of the BBH is  strongly affected by  the scalar
field. When the  scalar field is present,  the BBH merges  faster.
Notice that the ADM  mass is not  the main  cause of the  fast
merge. As  shown in Table~\ref{tab:3},  for cases $\phi_0=0.00048$
and $\phi_0=0.000959$, the ADM mass is  larger than in the GR case.
On  the other hand, when $\phi_0=0.000048$,  the  ADM masses  for
$a_2\rightarrow \infty$  and $a_2=2.61877$    are   smaller    and
equal    to   the    GR   case respectively. However, in both  cases
with non-vanishing scalar field, the BBH merges faster than in the
GR case (see Fig.~\ref{fig:13}).
%|--------------------------------------------------------------------|

\begin{figure}[tbp]
  \centering
  \includegraphics[width=0.5\textwidth]{\imgdir/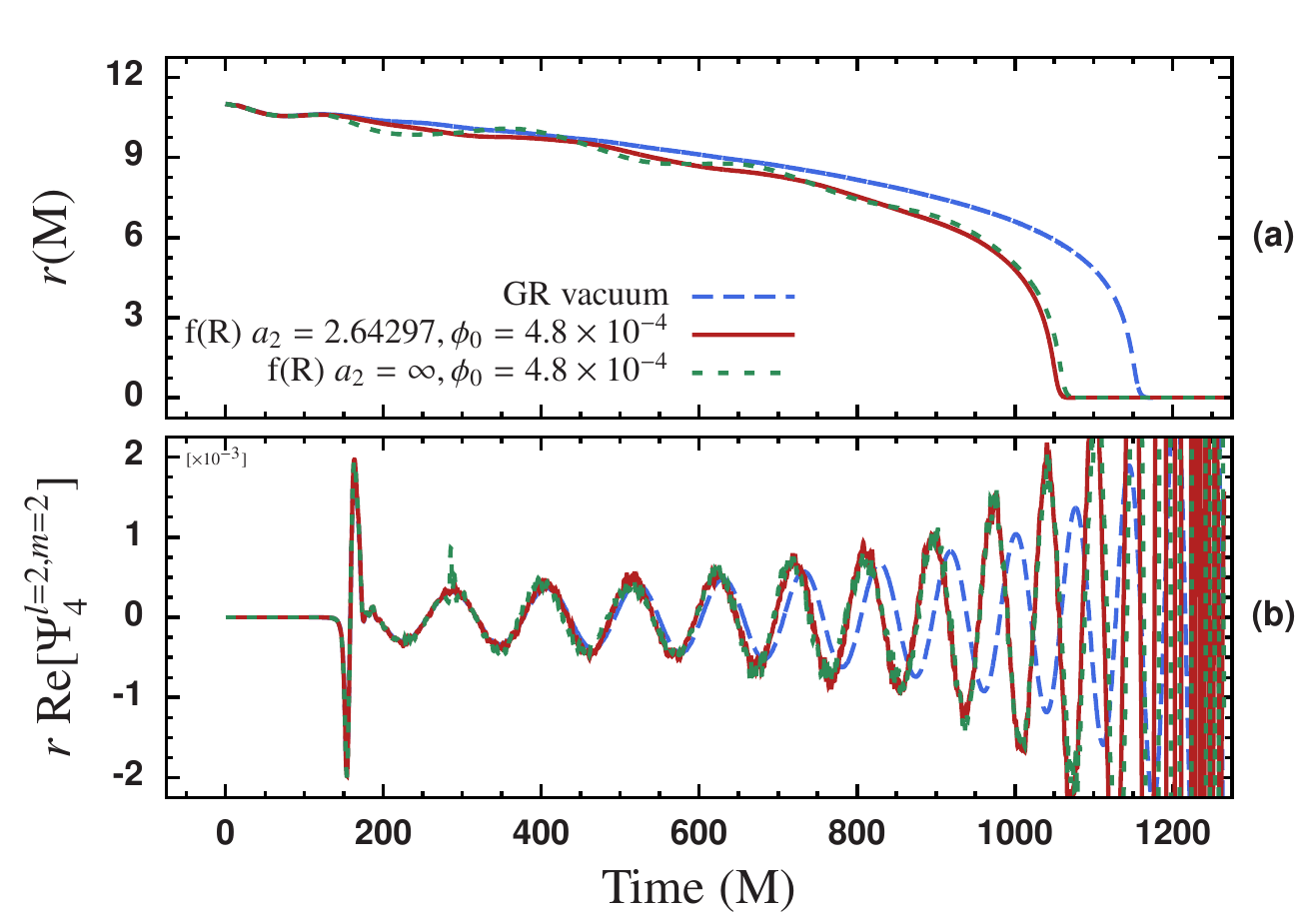}
  \caption{The  upper  panel~\bfa~  shows  the  coordinate  separation
    between the black holes.  The lower panel~\bfb~ shows the waveform
    ($\ell=2$,   $m=2$  mode).   The  \fR   effect   introduces  extra
    eccentricity to the BBH orbit.}\label{fig:14}
\end{figure}

%|--------------------------------------------------------------------|
For larger values of $\phi_0$, for example $0.00048$, the scalar field
increases the eccentricity of the BBH's orbit in addition to making it
merge faster.  This extra eccentricity depends on the parameter $a_2$.
When  $a_2$   is  big,  the  resulting  eccentricity   is  large  (see
Fig.~\ref{fig:14}-\bfa). In  addition, we observe that  the \fR effect
makes the BBH  merge faster in finite $a_2$ case than  in the free EKG
case.
%|--------------------------------------------------------------------|
Previously  in  Sec.~\ref{sec:dynamic-sf-bbh},  we  noticed  that  the
interaction between the  scalar field and the black  hole is weaker in
finite $a_2$  case than in the  free EKG case.  The  behavior shown in
Fig.~\ref{fig:14}-\bfa~ is consistent  with this conclusion.  When the
interaction  is  stronger,  it  introduces more  eccentricity  to  BBH
evolution.   More  eccentric  BBH  orbits produce  more  gravitational
radiation \cite{GolBru10}. Therefore, the mergers are faster.
%|--------------------------------------------------------------------|

%|--------------------------------------------------------------------|
Although the coordinate information is gauge dependent, it is possible
to verify a change in the eccentricity by looking at the gravitational
waves (see Fig.~\ref{fig:14}-\bfb). Notice  that the amplitude of the
gravitational radiation burst in finite  $a_2$ case is smaller than in
the free EKG case.
%|--------------------------------------------------------------------|
In Fig.~\ref{fig:10},  we can see  the change in the  eccentricity for
the case of $\phi_0=0.000959$.
%|--------------------------------------------------------------------|

%|--------------------------------------------------------------------|
%|--------------------------------------------------------------------|

%|--------------------------------------------------------------------|
So  far,  we  have  shown  that  small $\phi_0$  for  free  EKG  cases
introduces  more \fR  effects than  finite $a_2$  cases. On  the other
hand, large  $\phi_0$ for free  EKG cases introduces less  \fR effects
than finite  $a_2$ cases. It is  possible that the  nonlinear terms of
the finite $a_2$ cases are the cause of these differences.
%|--------------------------------------------------------------------|

%|--------------------------------------------------------------------|
Considering  the \fR  effect introduced  by the  scalar field,  we can
distinguish   the   parameter   $a_2$   through   gravitational   wave
detection. LIGO's main  BBH sources are black holes  with several tens
of solar mass. If $a_2$  is bigger than $10^{11}$m${}^2$, we expect to
be  able   to  distinguish  between   \fR  theory  and  GR,   via  the
gravitational  detection. On  the other  hand, LISA  (or  some similar
spacecraft  experiment)   can  distinguish  between  \fR   and  GR  if
$a_2>10^{17}$m${}^2$ \cite{BerGai11}.  All  together, the merger phase
of BBH collisions allows distinction between the theories, as proposed
by \cite{HeaBodHaa11}.
%|--------------------------------------------------------------------|

%|--------------------------------------------------------------------|
\subsubsection{Difference between \fR and other Einstein-Klein-Gordon models in GR}

%|--------------------------------------------------------------------|
\begin{figure*}[tbp]
\begin{tabular}{cc}
\includegraphics[width=0.5\textwidth]{\imgdir/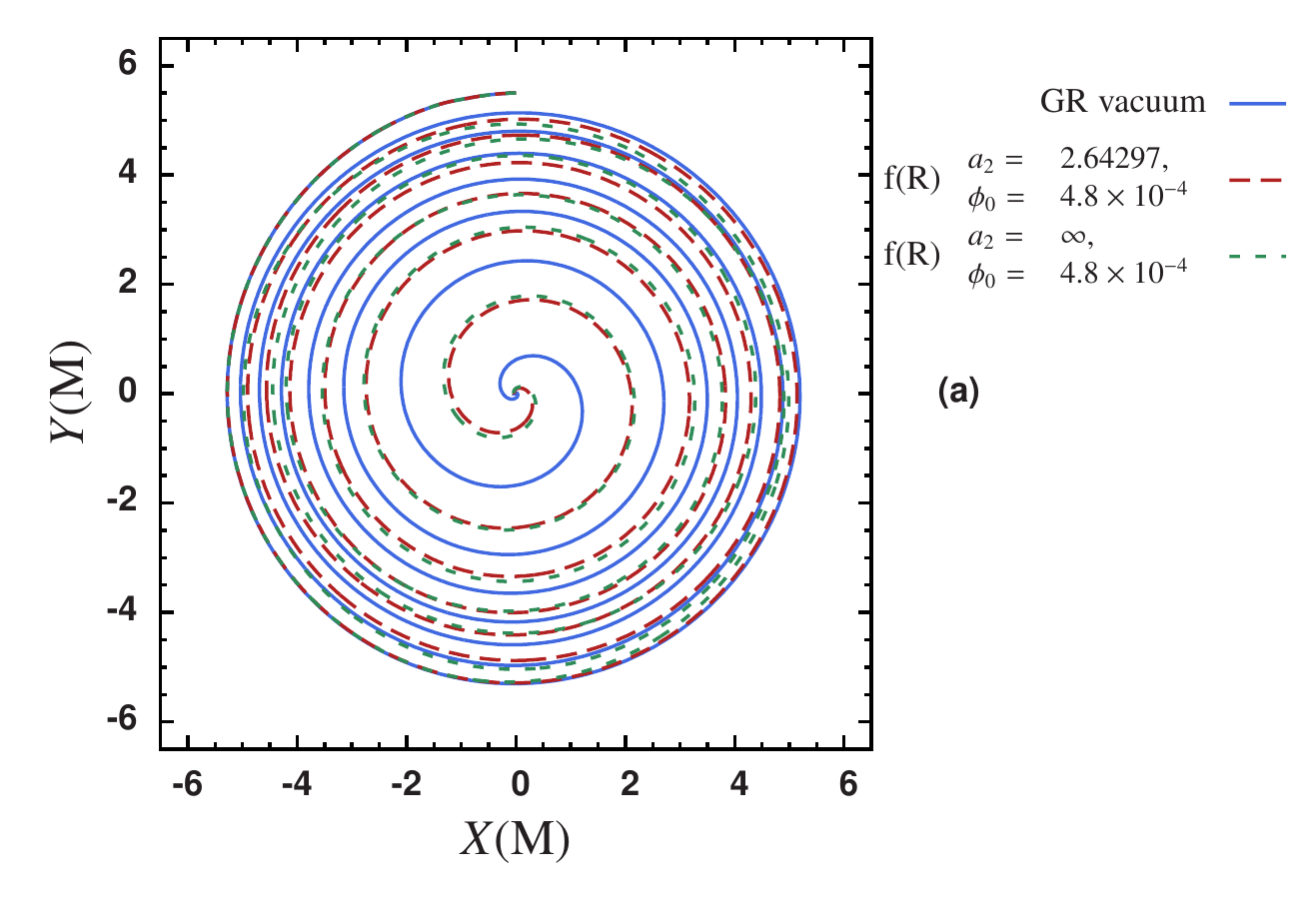}&
\includegraphics[width=0.5\textwidth]{\imgdir/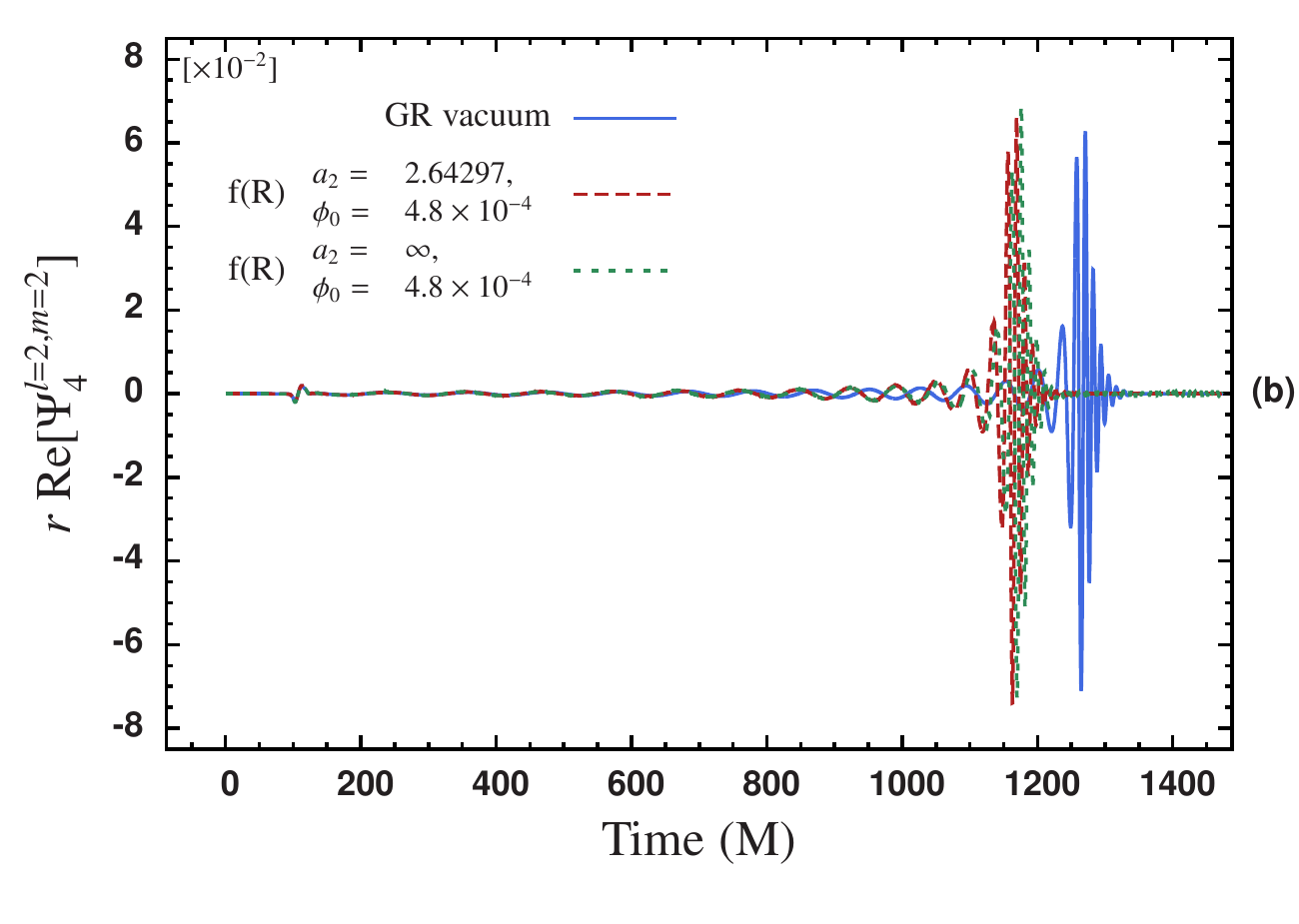}\\
\includegraphics[width=0.5\textwidth]{\imgdir/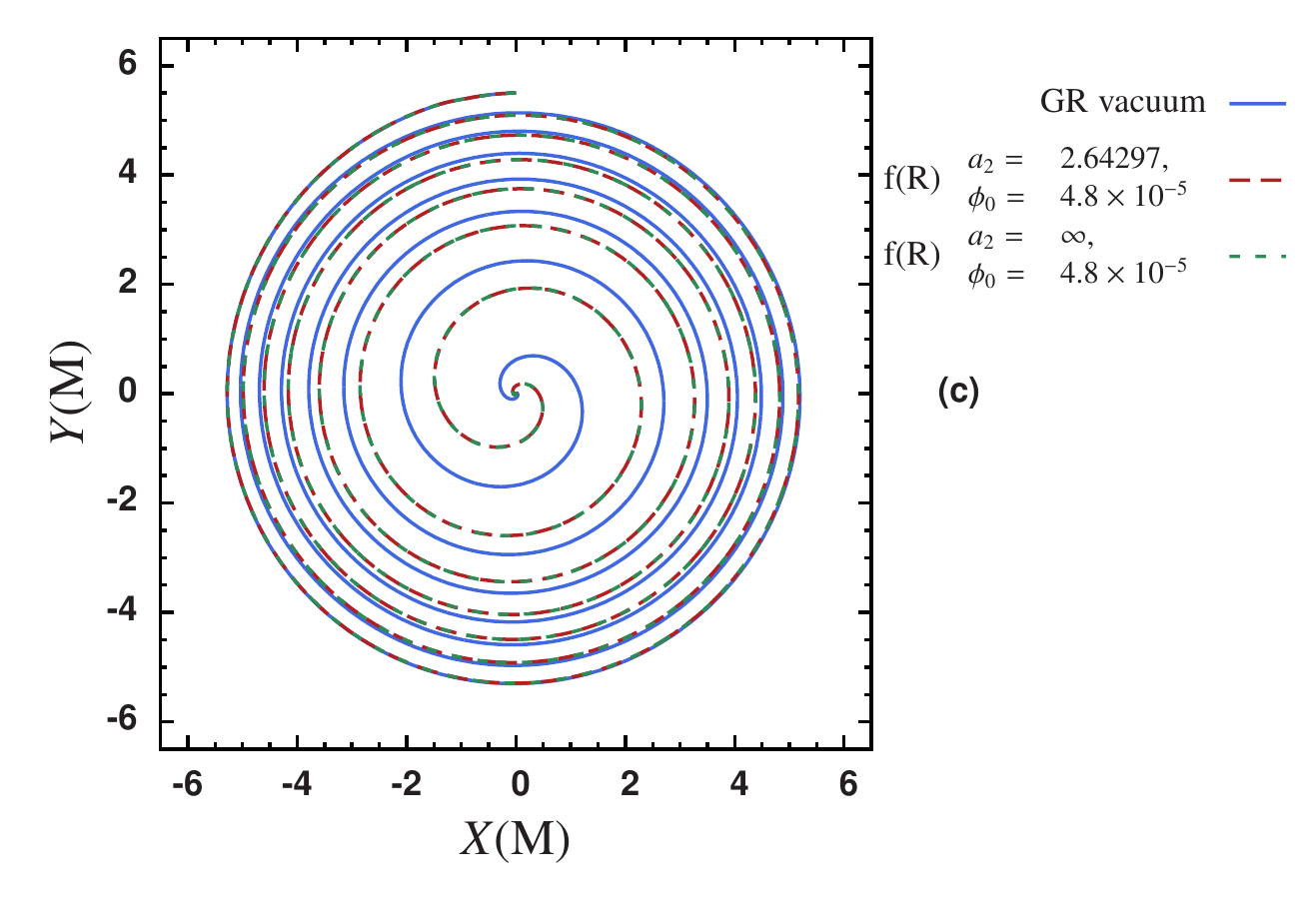}&
\includegraphics[width=0.5\textwidth]{\imgdir/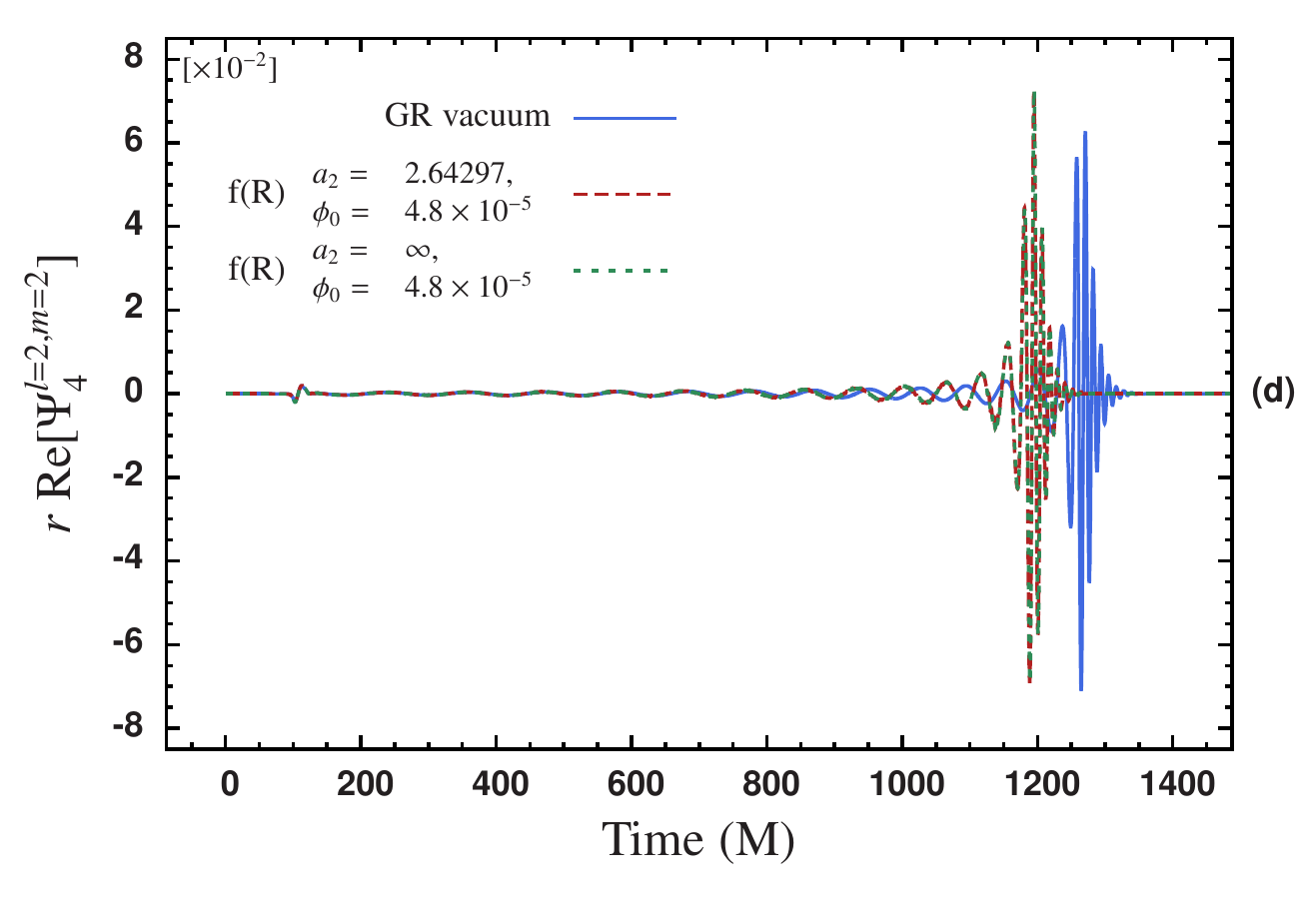}
\end{tabular}
\caption{ Trajectories and  waveforms.  Comparison between BBH mergers
  in GR, a  representative case of \fR and  the corresponding free EKG
  model. Panel~\bfa:  BBH trajectory for  vacuum GR (solid  line), \fR
  theory  (dashed  line) and  free  EKG  matter  model in  GR  (dotted
  line). We  show the trajectory  of one of  the two black  holes, the
  trajectory of the companion black  hole is symmetric with respect to
  the   $Z$   axis.   The   scalar   field   amplitude  parameter   is
  $\phi_0=4.8\times 10^{-4}$.  Panels \bfb: The corresponding waveform
  (real part  of $\Psi_4$, mode $\ell=2$, $m=2$).  Panel~\bfc: Same as
  in panel~\bfa~but with $\phi_0$ ten times smaller ($\phi_0=4.8\times
  10^{-5}$).    Panel~\bfd:  Corresponding   waveform  for   the  case
  $\phi_0=4.8\times 10^{-5}$. }\label{fig:15}
\end{figure*}
%|--------------------------------------------------------------------|
%|--------------------------------------------------------------------|
We  have seen above  that it  is possible  to distinguish  between \fR
theory and GR via the gravitational waves.  Astrophysical models often
include EKG  equations for the  description of certain  phenomena. For
example, there  are models of  dark matter which  use EKG in  the weak
field limit  \cite{AlcGuzMat02,BerGuz06, BerGuz06a, BarBerDeg11}.  One
example    of    relativistic   scalar    field    is   boson    stars
\cite{Bal99,BalBonGuz04,BalComShi98,     BalSeiSue97,    BalBonDau06}.
Therefore, it  is interesting to  ask if gravitational  wave detection
can be used  to distinguish BBH collisions in  \fR theory from another
system which also contains scalar fields.
%|--------------------------------------------------------------------|

%|--------------------------------------------------------------------|
In the  rest of this section,  we analyze the  differences between the
free EKG  system ($a_2 \rightarrow  \infty$) and the \fR  theory.  The
main difference between free EKG  and \fR theory is the nonlinear self
interactions,  present only  in \fR  theory.  If  the scalar  field is
strong, it  is easy to distinguish  between free EKG and  \fR.  If the
scalar  field is  weak, a  deeper analysis  is necessary  in  order to
distinguish  between the  theories. Our  quantitative  results support
this statement.
%|--------------------------------------------------------------------|

%|--------------------------------------------------------------------|
First    row   of    Fig.~\ref{fig:15}   shows    the    results   for
$\phi_0=0.00048$. Fig.~\ref{fig:15}-\bfa~ shows  the trajectory of one
of the components  of the binary (the companion  black hole trajectory
is  symmetric with  respect  to the  $X$  axis).  We  can see  several
crosses of the trajectories.  This indicates different fluctuations on
the  inspiral   rate.   This  results  from   the  extra  eccentricity
introduced by the  scalar field.  In Sec.~\ref{sec:dynamic-bbh-sf} and
Fig.~\ref{fig:14}, we saw that the  eccentricity is larger in the free
EKG system than in the representative case of \fR theory. In addition,
the BBH in  \fR theory merges faster than in  the free EKG. Therefore,
it is possible to distinguish between free EKG models and \fR theory.
%|--------------------------------------------------------------------|

\begin{figure}[tbp]
  \centering
  \includegraphics[width=85mm]{\imgdir/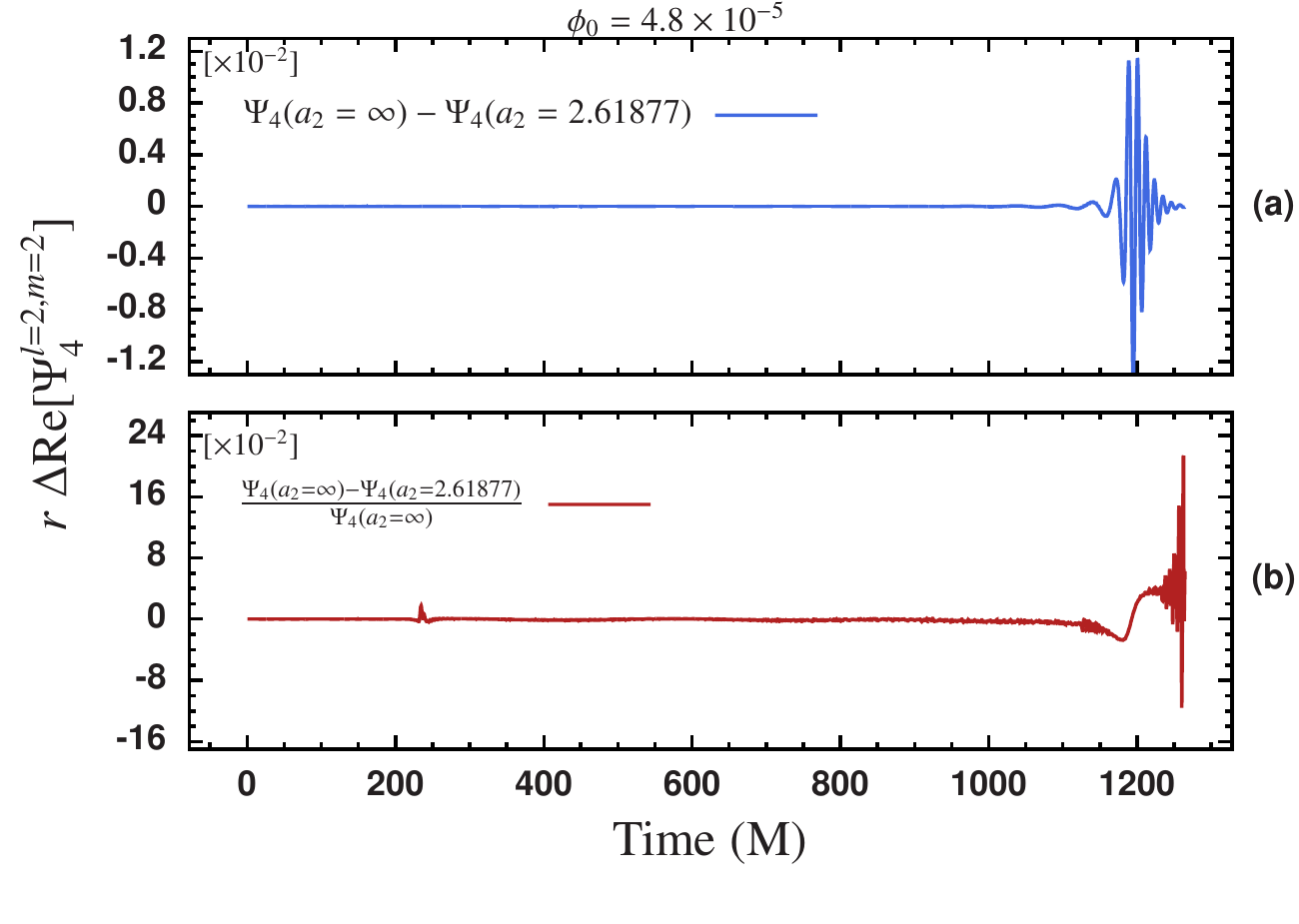}
\caption{Upper  panel  \bfa:  Difference  between  the  real  part  of
  $\Psi_{4}^{l=2,m=2}$  in  \fR theory  ($a_2=2.61877$)  and free  EKG
  model.  Lower  panel  \bfb:  Relative  difference  in  amplitude  of
  $\Psi_{4,22}$.}\label{fig:16}
\end{figure}

%|--------------------------------------------------------------------|
The second    row   of   Fig.~\ref{fig:15}    shows   the    results   for
$\phi_0=0.000048$  (the value is  ten times  smaller).  In  this case,
there are  no noticeable differences  between free EKG models  and \fR
theory.   This  is  consistent  with  our  assumption  that  the  self
interaction  becomes  weak  for  small  scalar  field.   However,  the
quantitative  difference of the  $\ell=2$, $m=2$  mode of  $\Psi_4$ is
significant  (see  Fig.~\ref{fig:16}-\bfa).   Moreover,  the  relative
difference      is     larger      than      ten     percent      (see
Fig.~\ref{fig:16}-\bfb). Once again, there  is a small peak at roughly
$\mathrm{time}=240$  M  in Fig.~\ref{fig:16}-\bfb.   The  peak is  the
result of a burst of  gravitational radiation produced by the free EKG
model, which is  absent in the \fR case  (see also Fig.~\ref{fig:10}).
We expect that  we will be able to  characterize the differences using
more  detailed  quantitative data  analysis  techniques.   We plan  to
present the results in a forthcoming paper.
%|--------------------------------------------------------------------|

%|--------------------------------------------------------------------|

%|--------------------------------------------------------------------|
\section{Discussion} \label{sec:discussion}
%|--------------------------------------------------------------------|

%|--------------------------------------------------------------------|
Extending the  work of \cite{BerGai11},  where the extreme  mass ratio
BBH systems were  considered to be the gravitational  wave sources for
LISA, we studied an equal mass BBH system.
%|--------------------------------------------------------------------|
In order  to simulate  BBH in \fR  theory with our  existing numerical
relativistic  code,  we  performed  transformations of  the  dynamical
equations of \fR  theory from the Jordan frame  to the Einstein frame.
In this way, we performed full numerical relativistic simulations.
%|--------------------------------------------------------------------|
The  main result  in \cite{BerGai11}  is that  the  gravitational wave
detection with LISA  can distinguish between \fR theory  and GR if the
parameter $|\mathbf{a_2}|>10^{17}$m${}^2$.  Our results imply that the
gravitational  wave   detection  with  LIGO   can  do  the   same  for
$|\mathbf{a_2}|>10^{11}$m${}^2$.
%|--------------------------------------------------------------------|

%|--------------------------------------------------------------------|
Mathematically, the dynamical equations  of \fR theory in the Einstein
frame  require  a scalar  field.   We  found  an interesting  dynamics
between this scalar  field and the BBH.  For  example, the BBH excites
the scalar  field for free  EKG cases ($a_2 \rightarrow  \infty$) near
the collision.   The scalar field  is constantly excited close  to the
BBH  for finite  $a_2$  cases.  Moreover,  the interaction  introduces
extra eccentricity to  the evolution of the BBH  orbit.  We found that
the  BBH eccentricity  is affected  by  the initial  parameter of  the
scalar  field $\phi_0$  depending on  the  value of  $a_2$. For  small
$\phi_0$,  the   excitation  of  the   BBH  orbit  is  larger   in  the
representative \fR case in comparison with the free EKG system. On the
other hand,  for larger values of  $\phi_0$ the excitation  of the BBH
orbit is smaller in the representative \fR case in comparison with the
free EKG system.
% |--------------------------------------------------------------------|

%|--------------------------------------------------------------------|
Using  gravitational waves, it  is possible  to distinguish  among \fR
theory, general relativity and a free Einstein-Klein-Gordon system. We
found that  the perturbation produced  by the scalar field  depends on
the  initial scalar field  configuration.  Specifically,  the waveform
exhibits a radiation burst  which depends quadratically on the initial
scalar  field amplitude.   The burst  is a  particular feature  of the
system which is useful when distinguishing between \fR and GR.
%|--------------------------------------------------------------------|
For  an  initial  amplitude  of scalar  field  $\phi_0=0.000048$,  the
relative difference  in the gravitational waveform  between \fR theory
and  the  free  EKG  model is  more  than  10\%.
%|--------------------------------------------------------------------|
Therefore,  gravitational wave  astronomy may  provide  the necessary
information  to rule  in or  rule out  some  alternative gravitational
theories.
%|--------------------------------------------------------------------|

\acknowledgments

%|--------------------------------------------------------------------|
It is a pleasure to thank David Hilditch, Ee Ling Ng, Todd Oliynyk and
Luis Torres  for valuable discussions and comments  on the manuscript.
This  work was  supported in  part by  ARC grant  DP1094582,  the NSFC
(No.~11005149, No.~11175019 and  No.~11205226), and China Postdoctoral
Science Foundation grant No. 2012M510563.
%|--------------------------------------------------------------------|

% -------------------------------------
%         BIBLIOGRAPHY
% -------------------------------------

\bibliography{refs,refs_extra}

% ----> END <----

\end{document}